\pdfoutput=1
\documentclass[notitlepage]{revtex4-1}
\usepackage[T1]{fontenc}
\usepackage{units}
\usepackage{textcomp}
\usepackage{bm}
\usepackage{amsmath}
\usepackage{amssymb}
\usepackage{graphicx}
\usepackage{esint}

\providecommand{\tabularnewline}{\\}

\begin{document}

\title{Single Molecule Spectroscopy of Monomeric LHCII: Experiment and Theory}

\author{Pavel Mal\'{y}$^{1,2},$ J. Michael Gruber$^{2}$, Rienk van
Grondelle$^{2}$, and Tom\'{a}\v{s} Man\v{c}al$^{1,2}$}

\affiliation{$^{1}$Institute of Physics, Charles University in Prague, Ke Karlovu
5, 12116 Prague, Czech Republic,$^{2}$Department of Physics and
Astronomy, Faculty of Sciences, Vrije Universiteit Amsterdam, De Boelelaan
1081, 1081HV Amsterdam, The Netherlands}
\begin{abstract}
We derive approximate equations of motion for excited state dynamics
of a multilevel open quantum system weakly interacting with light
to describe fluorescence detected single molecule spectra. Based on
the Frenkel exciton theory, we construct a model for the chlorophyll
part of the LHCII complex of higher plants and its interaction with
previously proposed excitation quencher in the form of the lutein
molecule Lut 1. The resulting description is valid over a broad range
of timescales relevant for single molecule spectroscopy, i.e. from
ps to minutes. Validity of these equations is demonstrated by comparing
simulations of ensemble and single-molecule spectra of monomeric LHCII
with experiments. Using a conformational change of the LHCII protein
as a switching mechanism, the intensity and spectral time traces of
individual LHCII complexes are simulated, and the experimental statistical
distributions are reproduced. Based on our model, it is shown that
with reasonable assumptions about its interaction with chlorophylls,
Lut 1 can act as an efficient fluorescence quencher in LHCII. 
\end{abstract}
\maketitle
\maketitle

\section*{Introduction}

Photosynthesis, arguably the most important photo-induced process
on Earth, converts the energy of light into its chemically/biologically
useful form. It is often argued that this conversion is remarkably
efficient. However, it has to be distinguished between the \emph{quantum}
efficiency, reaching almost unity\cite{BlankenshipBook,Wraight1974},
and light-to-chemical energy efficiency, which is significantly lower,
in the order of several percent\cite{Blankenship2011}. This is given
by an evolutionary pressure on the development of a robust photosynthetic
machinery optimized for survival rather than energy conversion efficiency.
Energy relaxation processes are integral part of the photosynthetic
function as they enable the energy transfer to proceed unidirectionally
\cite{VanAmerongen2000}, and provide protection of the photosynthetic
apparatus against harmful over-excitation. The photosynthetic machinery
of plants has developed a complex hierarchy of self-regulatory mechanisms
to avoid excess excitation or (when unavoidable) to dissipate it into
heat \cite{Ruban2012}. Starting from processes controlled on the
macroscopic level (e. g. orientation of leaves), over spontaneous
microscopic (cellular) events such as chloroplast movements, to truly
nano- and sub-nanoscopic mechanisms such as reorganization of antenna
complexes and direct regulation of energy transfer on the level of
small groups of interacting chromophores, plants actively react to
changing illumination conditions. The sub-nanoscopic processes, which
are the focus of the present study, operate in response to the increase
of $\Delta{\rm pH}$ across the thylakoid membrane. Such an increase
is an indicator of high illumination. Nowadays it is generally accepted
that carotenoids are involved in these energy dissipation processes.
The precise molecular mechanism is, however, still subject of discussion\cite{Holt2005,Ruban2007,Avenson2008,Bode2009,Berera2010,Muller2010,Fuciman2012,Staleva2015}.
It is likely that different mechanisms evolved in different classes
of organisms and/or that several mechanisms operate at once. 

Most of our current knowledge about the early processes of photosynthesis
was obtained by ultrafast spectroscopy. While conventional bulk spectroscopies
are extremely useful in following ultrafast photo-induced events in
photosynthesis, whenever structural inhomogeneity of the sample is
involved, the information obtained from these spectroscopies becomes
obscured by an inevitable ensemble averaging. Some well established
spectroscopic methods, such as hole-burning \cite{Haarer1990,Johnson1989,Creemers1999},
and some modern multidimensional methods, such as coherent two-dimensional
spectroscopy \cite{Jonas2003}, provide certain access to the homogeneous
properties. Information on the behaviour of individual molecules has
been, however, available only since the advent of single-molecule
spectroscopy (SMS) \cite{Moerner1989}. 

The system studied in this work, the light harvesting complex II
(LHCII) of higher plants, is the plants' major light-harvesting antenna
containing almost half of all the chlorophyll in the chloroplast.
Correspondingly, most of the light absorption and subsequent energy
transfer processes in plants and algae occur in this complex. The
LHCII antenna occurs naturally in a trimeric form and its main function
is to deliver excitation energy to the nearby photosystem II (PSII).
Given the major role of LHCII in light-harvesting and energy transfer,
it is not surprising that it is also implicated in participating in
regulated energy dissipation, the so-called non-photochemical quenching
(NPQ) \cite{Ruban2007,Bode2009,Schlau-Cohen2015}. The crystal structure
of the complex \cite{Liu2004} enables us efficient theoretical modeling
of the complex's spectroscopic properties using its chromophores (chlorophylls
(Chl) and carotenoids) as the model units. The model parameters such
as electronic coupling are greatly constrained by the known mutual
orientation and distances. Existence of a large body of previous measurements
together with theoretical attempts to fit this whole body of data
to a single model \cite{Novoderezhkin2004,Novoderezhkin2005a,Novoderezhkin2011}
gives a great confidence in extending the modeling towards single
molecular experiments.

In recent years, single-molecule spectroscopy (SMS) experiments on
several photosynthetic antenna complexes including the LHCII were
performed. Fluorescence spectral peak distributions \cite{Kruger2010},
spectral diffusion \cite{Tietz2001}, fluorescence intensity distributions
\cite{Kruger2011,Kruger2012} and time traces \cite{Kruger2011a}
were obtained from these measurements. In many cases also fluorescence
intermittency (blinking) was observed, and it was conjectured that
in the case of LHCII the process behind the fluorescence intermittency
plays a role in NPQ \cite{Kruger2011}.

As far as the theory of the SMS of LHCII is concerned, the ensemble-averaged
spectra and also the peak distributions of LHCII trimers can be successfully
explained by the disordered excitonic model \cite{Kruger2010}. In
general, the excitonic model was successfully applied in the past
on fs to ps timescale, and it represents an indispensable tool in
analyzing ultrafast spectroscopic experiments on molecular aggregates
and in particular on photosynthetic antennae\cite{VanAmerongen2000}.
Its application to longer time scales of seconds and minutes is conditioned
by the assumption of a certain separation of time-scales. Over the
course of the excitation-emission cycle (nanoseconds), individual
chromophores of a complex are assumed to be found in fixed spatial
arrangements and experiencing an environment described by a fixed
set of parameters. The emission spectrum of a complex in such a fixed
spatial arrangement is predicted by the excitonic model which gives
the population distribution of the excited state manifold and the
probability of emission at the corresponding transition frequencies.
The spectrum of exciting light matters only to the extent to which
the excited states reached at a given excitation wavelength are connected
to the final state by some relaxation pathways. Once the pathways
are available, the final state is given irrespective of the initial
state after absorption of light. Because only the final distribution
of excited state population matters, the changes (switches) of the
spatial arrangement or environmental conditions occurring on the sub-nanosecond
time scale are only observed as sudden changes (with respect to the
nanosecond fluorescence time scale) of the fluorescence spectra. Despite
the fact that the experiment we analyze in this work does not provide
more insight into the sub-nanosecond dynamics of individual complexes
than previous works, we nevertheless formulate the theory consistently
in such a way that it enables the description of such processes. This
is done in order to highlight the existence of a less studied type
of processes on an intermediate time scale and to stress the need
to search for experimental techniques which can cover the range of
timescales from femtoseconds to nanoseconds in single molecular spectroscopy.
We thus provide theoretical techniques to treat these future experiments.
We have recently reported a progress towards measuring processes on
the order of hundreds of femtoseconds in single light-harvesting complexes
\cite{Maly2015}. 

The paper is organized as follows: In Section \ref{sec:Excited-State-Population}
we first discuss the Frenkel exciton model as a basis for the formulation
of equations of motion for the populations of the excited states of
chromophore aggregates with strong interchromophore couplings. Then, in Section \ref{sub:Excited-State-Dynamics}, we
introduce equations of motion for excitonic populations valid over
a broad (from ps to min) timescale range. We discuss their generality
and the range of validity. It is argued that these equations provide
an ideal means for the description of the SMS experiments. In Section
\ref{sec:LHCII} the results of Section \ref{sub:Excited-State-Dynamics} are applied
to LHCII photosynthetic antenna complexes. All results are compared
to the experiment. It is shown that our equations give correct fluorescence
spectrum and peak statistics, i. e. appropriate steady-state population,
under typical SMS experiment condition. Then we model the intensity
traces, while the switching behaviour is included by incorporating
one particular previously proposed NPQ mechanism, namely energy transfer
to lutein Lut 1. The switching between on and off states is controlled
by a 2-level model, where the switching causes a change of the Chl
a612 - Lut 1 coupling. It is shown that using realistic parameters
we are able to reproduce the experimental results. The details of
our energy relaxation theory and of the stochastic model of switching
are given in Supporting Information (SI).

\section{Results}

\subsection{Frenkel-Exciton model\label{sec:Excited-State-Population}}

In the usual SMS experiments, time-resolved (on the times scale of
10 ms to s) fluorescence of the studied molecules is observed. If
the excited state life time of the studied chromophores is sufficiently
long (nanoseconds in the case of chlorophylls studied here), the expected
fluorescence spectra can be calculated from the steady state, quasi-equilibriated
populations of the excited electronic eigenstates of the molecular
system, assuming canonical equilibrium. Depending on the strengths
of the chromophore-chromophore resonance interaction and the system-bath
couplings, the eigenstates can be approximated either by the excited
eigenstates of the electronic subsystem or the excited state of the
individual chromophores forming the aggregate. In the present paper,
we treat both cases within one formalism provided by the framework
of the Frenkel exciton model. 

The Frenkel exciton model provides an excellent tool for the treatment
of pigment-protein complexes on femto- and pico-second time scales
\cite{VanAmerongen2000}. The basic notion of the Frenkel exciton
model is the one of localized excited states of the chromophores.
These states have negligible wavefunction overlap with neighboring
excited states, and they can therefore be assumed orthogonal to each
other, forming a suitable basis for the aggregate Hilbert space. In
the treatment of linear absorption and fluorescence experiments, only
singly excited collective states of the molecular aggregate have to
be included. The localized excited states can thus be denoted as $|i\rangle=|e_{i}\rangle\prod_{n\neq i}|g_{n}\rangle,$where
$|g_{i}\rangle$ and $|e_{i}\rangle$ are the electronic ground and
excited states of the chromophore denoted by index $i$, respectively.
The states $|i\rangle$ form a complete Hilbert subspace for the case
that exactly one molecule of the aggregate is excited. The molecular
system Hamiltonian, $H_{S}$, is however rarely diagonal in the basis
of the states $|i\rangle$ and resonance couplings $J_{ij}$ between
excited electronic states $|i\rangle$ and $|j\rangle$ occur in all
situations interesting for light-harvesting. In the absence of a protein
environment, the light would resonantly excite eigenstates of the
system Hamiltonian. In all realistic cases, the interaction of the
system with its environment co-determines the excitation frequencies,
and prescribes thus a ``preferred basis'' of electronic states in
which it is the most advantageous to formulate the energy transfer
theory.

In the case of the resonance interaction $J_{ij}$ exceeding the typical
reorganization energy $\lambda$ of the protein bath of the chromophores
$i$ and $j$, ($J_{ij}>\lambda)$, we include $J_{ij}$ explicitly
into the system Hamiltonian. The Hamiltonian is diagonalized to obtain
electronic eigenstates, and the effect of the protein bath is included
via perturbation theory yielding a Redfield type relaxation tensor.
In the opposite case $(J_{ij}<\lambda)$ we neglect the resonance
coupling in the system Hamiltonian, and its effects are included perturbatively
yielding F\"{o}rster type energy transfer rates between localized
excited states \cite{VanAmerongen2000,MayKuhn,MancalBook}.

\subsection{Excited State Dynamics across Time-Scales\label{sub:Excited-State-Dynamics}}

Let us first focus on deriving a closed set of equations for the excited
state populations. Extensive work was done in the last years on developing
methods to accurately describe the system dynamics following an ultrafast
excitation by external light. The traditional Redfield and F\"{o}rster
approaches were superseded by more accurate (exact in some cases)
methods, such as HEOM \cite{Ishizaki2009,Hein2012,Wilkins2015}, TEDOPA
\cite{Prior2010} and other methods \cite{Shi2004,Stockburger2002,Xu2005,Huo2010,Schroter2014}.
These methods bring unprecedented accuracy at an increased numerical
cost. It seems, however, that for analyzing many of the recent experiments,
it is still possible to rely on the traditional tools, as they capture
the physics of the problem (and often even the quantitative aspects
of the problem) very well \cite{Wilkins2015}.

Commonly, equations of motion for some relevant degrees of freedom
(DOF), electronic states in our case, are derived by reducing the
Liouville - von Neumann equation for the total density matrix to an
equation for the so-called reduced density matrix (RDM). These equations
describe time evolution of a molecular system for a fixed configuration
of the protein environment (assuming fast fluctuations around this
fixed configuration), and they are therefore suitable for the description
of ultrafast laser experiments. However, on the timescale relevant
in SMS (up to tens of seconds), usage of these equations is actually
not appropriate. First of all, some transient effect at the short
times affect even the long time properties of the system and the steady
state population dynamics, and second, slow changes in the protein
environment can entirely change the energy landscape including the
case that one has to change the theoretical limit (localized states,
delocalized states) in which one works. The latter case is especially
difficult to treat and requires to go beyond the traditional master
equation approaches which we apply here. In this work we therefore
concentrate on the extension of the validity of the master equation
approach towards long times under the assumption that the dependence
of the Hamiltonian of the system on time is negligible within one
absorption emission time scale (nanoseconds). As for the transient
effects at short time, when dealing with fast dephasing of optical
coherences, short time transient effects are responsible for the absorption
lineshape. This aspect of the transient effects will be taken into
account in full.

Spontaneous emission of photons by chlorophylls occurs with a nanosecond
life-time. Another class of transient effects, namely dynamic electronic
coherence due to excitation by light, is therefore also unimportant
and its treatment can be avoided. We therefore derive approximative
equations for the populations dynamics only, with the validity in
the range from picoseconds to tens of seconds. 

Initially, our Hamiltonian consists of the system, bath and system-bath
interaction terms, and we formally assume some total density matrix
$W(t)$ which follows the Liouville--von Neumann equation. There are
several methods how to arrive to a master equation (in a convolution-less
form) RDM $\rho(t)=Tr_{B}W(t)$ \cite{Fain2000,MancalBook}. In general,
such equation reads:

\begin{equation}
\frac{\partial}{\partial t}\rho(t)=-\frac{i}{\hbar}[H(t,t_{0}),\rho(t)]-{\cal R}(t,t_{0})\rho(t),\label{eq:equation_of_motion}
\end{equation}

\begin{equation}
H(t,t_{0})=H_{S}(t_{0})-\mu E(t).\label{eq:total_Ham}
\end{equation}

The total Hamiltonian $H$ consists of the bath renormalized system
Hamiltonian $H_{S}(t_{0})$ and the system-light interaction, which
is (in the dipole approximation) given by the dipole moment operator
$\mu$ and electric field $E(t)$. The bath is completely eliminated
in the reduced description, and its effects are represented by the
relaxation tensor ${\cal R}(t,t_{0})$. It is important to note that
in Eq. (\ref{eq:equation_of_motion}), the relaxation tensor ${\cal R}(t,t_{0})$
explicitly depends on some initial time $t_{0}$, in which the initial
condition for the propagation is set, or more precisely, the relaxation
tensor depends on the quantity $t-t_{0}$. An exact relaxation tensor
would also depend on $E(t)$. In general, the choice of $t_{0}$ is
arbitrary, and we should choose it such that the temporal profile
of the excitation field is non-zero for $t>t_{0}$ only. However,
in most practical theoretical approaches, the evolution of the bath
due to excitation of the system is reflected in the relaxation tensor
${\cal R}(t,t_{0})$ by the bath correlation function, which decays
rather quickly. By choosing $t_{0}$ sufficiently small, we could
always make such a relaxation tensor constant. This is obviously an
artifact of the approximations used. For a smooth envelope of the
external field changing on the same or slower timescale than the correlation
function of the bath, there is no good choice of $t_{0}$. The relaxation
tensor always becomes constant too early. Luckily, as we will see
below, we do not need to account for the external field $E(t)$ beyond
first order perturbation theory, and in this regime, there is a natural
choice of time $t_{0}$ which enables us to correctly account for
the transient time dependence of the relaxation tensor even for a
steady state externally driven by light. It is important to note,
that the relaxation tensor ${\cal R}(t,t_{0})$ is completely abstract
up to now. It can represent some exact relaxation superoperator, or
a result of perturbation theory with respect to some parameters, such
as the Redfield or F\"{o}rster tensors.

Let us now formulate the equations of motion for the excited state
evolution in the linear regime of the system's interaction with the
electric field. The validity of the linear regime has been discussed
e.g. in Refs. \cite{Jiang1991,Mancal2010,Brumer2012a,Chenu2014},
and it is the same as the validity of the third order response theory
for non-linear laser spectroscopy. We will also use the secular approximation
(equations of motion for populations $\rho_{ii}$ and coherences $\rho_{ij}$,
$i\neq j$ are assumed independent), although this we do only to simplify
the resulting equations. Secular approximation could be avoided if
one so wishes, at the cost of treating the full DM. We write explicitly
the elements of the RDM of Eq. (\ref{eq:equation_of_motion}) in which
we keep the double time dependence of the Hamiltonian and the relaxation
tensor. We get the following set of coupled equations: 
\begin{equation}
\frac{\partial\rho_{ii}(t)}{\partial t}=\sum_{j}k_{ij}(t,t_{0})\rho_{jj}(t)-\Gamma_{i}(t,t_{0})\rho_{ii}(t)+(i\mu_{i0}\rho_{0i}(t)E(t)+c.c.),\label{eq:populations eq}
\end{equation}
\begin{equation}
\frac{\partial\rho_{i0}(t)}{\partial t}=-i\omega_{i0}(t_{0})\rho_{i0}-\gamma_{i}(t,t_{0})\rho_{i0}(t)+i\mu_{i0}(\rho_{00}-\rho_{ii}(t))E(t).\label{eq:coherences eq}
\end{equation}
Here, the $k_{ij}(t,t_{0})\equiv{\cal R}_{iijj}(t,t_{0})$ are transfer
rates between populations (from $j$ to $i$), $\Gamma_{i}(t,t_{0})\equiv\tilde{\Gamma}_{i}+{\cal R}_{iiii}(t,t_{0})$
is the population relaxation rate of the state $i$, including the
rate of radiative depopulation $\tilde{\Gamma}_{i}$ (we assume it
is constant), and $\gamma_{i}(t,t_{0})\equiv{\cal R}_{i0i0}(t,t_{0})$
is the optical coherence dephasing rate. 

The equation of motion for the optical coherences, Eq. (\ref{eq:coherences eq}),
can be solved by introducing the interaction picture using the evolution
operator element ${\cal U}_{i0i0}(t,t^{\prime})$, and integrating
Eq. (\ref{eq:coherences eq}) in the interaction picture. By returning
back to the Schr\"{o}dinger picture, we obtain the actual field-induced
and field driven optical coherence element in the form:
\begin{equation}
\rho_{i0}(t)=i\int_{0}^{t-t_{0}}dt^{\prime}{\cal U}_{i0i0}(t,t-t^{\prime})\mu_{i0}[\rho_{00}(t-t^{\prime})-\rho_{ii}(t-t^{\prime})]E(t-t^{\prime}).\label{eq:coherence_optical}
\end{equation}
We can see that Eq. (\ref{eq:coherence_optical}) is actually not
a solution of Eq. (\ref{eq:coherences eq}). In solving Eq. (\ref{eq:coherences eq})
numerically, the dephasing rate would quickly become constant for
$t\gg t_{0}$ and the transient effects for $t-t_{0}\approx0$ would
be lost completely. However, in Eq. (\ref{eq:coherence_optical})
these effects are properly taken into account. The discrepancy between
Eq. (\ref{eq:coherences eq}) and Eq. (\ref{eq:coherence_optical})
is due to a different treatment of the bath. In a response function
approach, of which the linearization of the full equations of motion
with respect to the field is a variant, it is in general possible
to account for the bath in a more consistent way than in master equations.
In ordinary master equations, the bath is correctly described at $t_{0}$,
and the description of its subsequent evolution after $t_{0}$ is
extremely limited. An example of such limitation is discussed e.g.
in Ref. \cite{Mancal2012}.

In most biological energy transferring systems, the pure dephasing
is much faster (hundreds of fs) than the population relaxation (units
and tens of ps). We can therefore assume that the populations remain
constant during the integration in Eq. (\ref{eq:coherence_optical}).
Moreover, it is reasonable to set the upper limit of the integration
in Eq. (\ref{eq:coherence_optical}) to infinity by sending $t_{0}\rightarrow-\infty$.
We can expect that the simultaneous action of pure dephasing and external
driving by a field with a slowly varying envelop creates a steady
state optical coherence. The dependence on $t_{0}$ in the upper limit
of the integral describes a transient evolution of the optical coherence
after switching on the interaction with the field. Now that the short
time time-dependent nature of the dephasing rates is taken into account
correctly, we can set $t_{0}\rightarrow-\infty$ and write: 

\begin{equation}
\rho_{i0}(t)=i\int_{0}^{\infty}dt^{\prime}{\cal U}_{i0i0}(t,t-t^{\prime})E(t-t^{\prime})\mu_{i0}[(\rho_{00}(t)-\rho_{ii}(t)].\label{eq:coherence FT}
\end{equation}
It is important to note that now $t$ is a global time which can run
through the whole minutes long SMS experiment. The evolution operator
changes on an ultrafast time scale, but this timescale is scanned
in the integration over the variable $t^{\prime}$. The properties
of the Hamiltonian $H_{S}(t)$ change on a very slow time scale (with
respect to optical dephasing), and so does the evolution operator
element ${\cal U}_{i0i0}(t,t-t^{\prime})$. Unlike the Eq. (\ref{eq:coherences eq}),
which is valid for $t-t_{0}$ small, Eq. (\ref{eq:coherence FT})
is valid for all times. 

The purpose of deriving Eq. (\ref{eq:coherence FT}) was to insert
it eventually into Eq. (\ref{eq:populations eq}). Also here we face
the problem of transient time dependence of the rates. However, because
populations change much more slowly, these effects are not as important
as in the case of coherences. They can be, however, treated rigorously,
even including bath memory effects between evolutions by Eq. (\ref{eq:coherences eq})
and Eq. (\ref{eq:populations eq}), as we have shown elsewhere \cite{Olsina2012}.
In this work, we will assume the energy transfer and relaxation rates
not to depend on the difference $t-t_{0}$, although they may depend
weakly on the time $t$, i. e. $k_{ij}(t,t_{0})=k_{ij}(t)$ and $\Gamma_{i}(t,t_{0})=\Gamma_{i}(t)$.
We are ready now to insert Eq. (\ref{eq:coherence FT}) for optical
coherences into the equation for populations, Eq. (\ref{eq:populations eq}).
We obtain 

\begin{equation}
\frac{\partial P_{i}(t)}{\partial t}=\sum_{j}k_{ij}(t)P_{j}(t)-\Gamma_{i}(t)P_{i}(t)+|\mu_{i0}|^{2}(P_{0}(t)-P_{i}(t))\cdot2{\rm Re}\int_{0}^{\infty}dt^{\prime}E(t)E(t-t^{\prime}){\cal U}_{i0i0}(t,t-t^{\prime}).\label{eq:driven_pop_evol}
\end{equation}
Here, we defined $P_{i}(t)=\rho_{ii}(t)$, and we used the fact that
$\mu_{i0}$ and even ${\cal U}_{i0i0}(t,t^{\prime})$ are ordinary
c-numbers. 

In Eq. (\ref{eq:driven_pop_evol}), the populations are driven by
a second order field term. We have treated the field classically so
far. If we did that quantum mechanically, we would now have to trace
over the field DOF in order to obtain reduced equations of motion
for the electronic state populations only. The term $E(t)E(t-t^{\prime})$
would thus be replaced by $\left\langle E(t)E(t-t^{\prime})\right\rangle $
which can be interpreted as a quantum mechanical expectation value.
The latter expression has the form of two-time correlation function
of the electric fields of the light and its Fourier transform is the
power spectrum of the light \cite{Jiang1991,Brumer2012a},

\begin{equation}
\left\langle E(t)E(t-\tau)\right\rangle =\int_{0}^{\infty}d\omega W(\omega)e^{i\omega\tau}.\label{eq:power_spectrum}
\end{equation}
Now, inserting Eq. (\ref{eq:power_spectrum}) into Eq. (\ref{eq:driven_pop_evol}),
switching the order of integrations, and using the definition of absorption
lineshape of the i-th excitonic transitions (see e.g. \cite{MukamelBook}) 

\begin{equation}
\chi_{i}(\omega;t)=|\mu_{i0}|^{2}\int_{0}^{\infty}dt^{\prime}e^{i\omega t^{\prime}}{\cal U}_{i0i0}(t,t-t^{\prime}),
\end{equation}
we arrive at

\begin{equation}
\frac{\partial P_{i}(t)}{\partial t}=\sum_{j}k_{ij}(t)P_{j}(t)-\Gamma_{i}(t)P_{i}(t)+\left[P_{0}(t)-P_{i}(t)\right]\int d\omega W(\omega)\chi_{i}(\omega;t).\label{eq:PopEvolution}
\end{equation}

Our effort has yielded a closed set of equations for excitonic populations
only. The population changes are given by the transfer rates between
electronic levels, population quenching and a source terms expressed
as an overlap of the excitonic spectra with the light spectrum. All
quantities are in principle dependent on time, most importantly the
excitonic absorption spectrum and all rates can weakly depend on time
to simulate slow changes of the protein and chromophore configurations.
Also the light spectrum can be considered time-dependent by generalizing
the Wiener-Khintchine theorem for the instaneous power spectrum $W(\omega,t)$\cite{Page1952,Jiang1991}.
The changes can be faster than the time resolution of the SMS experiment,
but they have to be slower than the dephasing, or even energy transfer
dynamics. 

We should also note here that if we wish to stay strictly in the linear
regime, the saturation term $(P_{0}-P_{i})$ in Eq. (\ref{eq:PopEvolution})
should be very close to one. In fact, we can set it equal to one as
a reasonable approximation with the same validity as the linear approximation
for fields.

\subsection{Spectroscopy of LHCII Complex\label{sec:LHCII}}

\subsubsection{Excitonic Model for Bulk and Single Molecule Spectra\label{sub:Excitonic-Model-for}}

According to crystalographic studies, LHCII complex consists of three
monomeric units, each containing 14 chlorophylls and for carotenoids:
two luteins, neoxanthin and a carotenoid of the xanthophyll cycle\cite{Liu2004}.
The experiments described in this section were performed on monomeric
LHCII complexes. In accord with the experiment, we focus on one such
a monomeric unit in our model. We treat the LHCII monomer as a strongly
coupled systems of chromophores, weakly coupled to the bath and weakly
interacting with light. Because only absorption and fluorescence are
measured, we do not attempt to fit the site energies in our simulations
from scratch, as the fitting of this limited set of experiments would
not be unique. Instead, we take the pigment transition energies from
Ref. \cite{Novoderezhkin2005a}, where both LHCII trimers and monomers
where treated. The coupling energies between the pigments were calculated
in the dipole-dipole approximation, and the dipole orientations were
taken from the crystal structure using an effective dipole strengths
of 3.4 D for Chl b and 4.0 D for Chl a. The bath is described by means
of a spectral density obtained by fluorescence line narrowing experiment
(FLN) (see Ref. \cite{Novoderezhkin2004,Peterman1997}). Excitonic
absorption and fluorescence lineshapes are calculated by means of
the second order cumulant expansion (see Ref. \cite{MukamelBook}),
and the population transfer rates are calculated by the multilevel
Redfield theory (see Ref. \cite{MayKuhn}). For comparison, we also
calculated the rates by Modified Redfield theory, Ref.\cite{Novoderezhkin2004},
and we concluded that the results remain essentially the same. The
population relaxation rates of chlorophylls were
taken to be 3 ns in accord with the experiment \cite{Nordlund1981}.
The equations of motion, Eq. (\ref{eq:PopEvolution}), allow us to
use light with any spectral composition. In the experiment described
in this paper, we use spectrally narrow (laser) illumination at 630
nm. For details on the calculations see SI. 

\begin{figure}[h]
\includegraphics{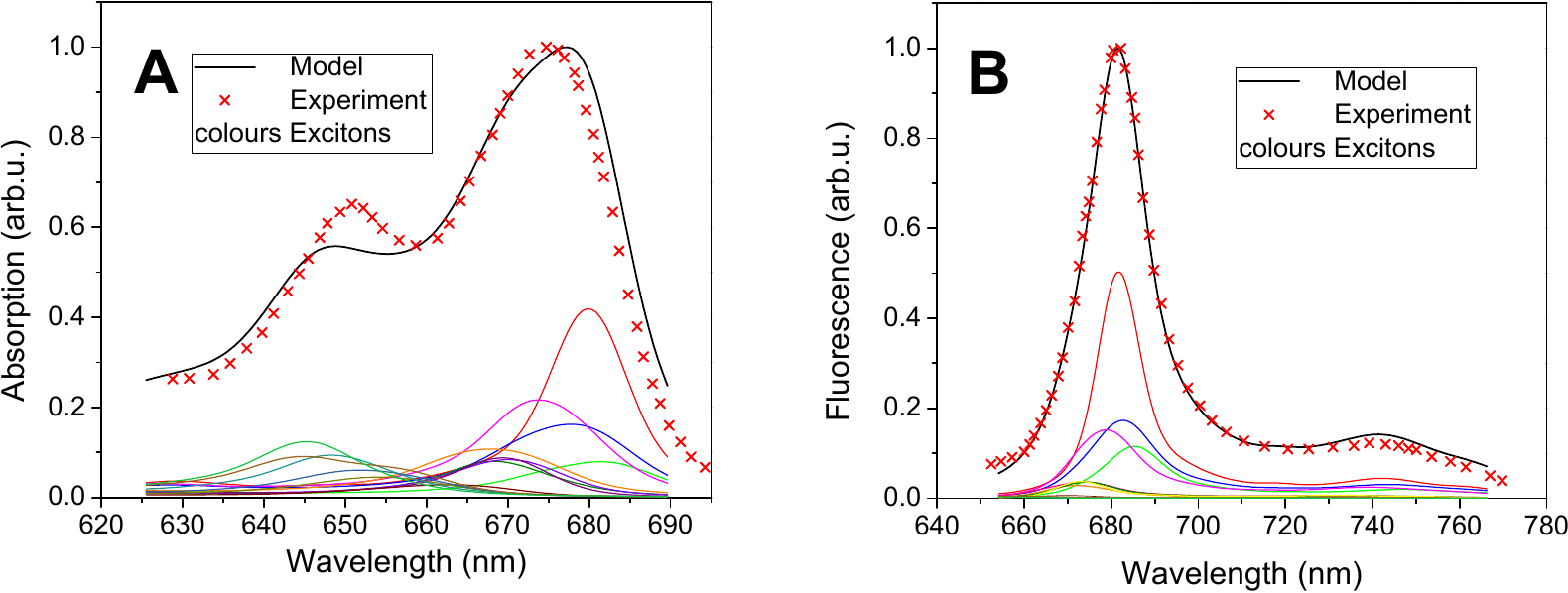}

\caption{\label{fig:Excitonic-model}Bulk (A) Absorption and (B) Fluorescence
spectrum of the Qy band of LHCII monomers. The points are experimental
values taken from Ref. \cite{Kruger2010}, the lines are calculated
by our exciton model. The coloured lines are individual excitonic
contributions, the black line is the overal spectrum. }
\end{figure}

In Figs. \ref{fig:Excitonic-model}A and \ref{fig:Excitonic-model}B
we present the calculated bulk absorption and fluorescence spectra
of LHCII monomers at 5\textdegree{} C compared to experimental values
taken from \cite{Kruger2010}. We note here that the bulk spectra
of LHCII in monomeric and trimeric form are practically identical,
see Fig. S1 in SI for comparison. In calculations, the spectra were
averaged over a Gaussian disorder of site energies with full width
at half maximum (FWHM) of 110 cm$^{-1}$. Although the blue Chl b shoulder
is not perfectly reproduced, the agreement between measured and theoretical
absorption is good in the region of our excitation, and the fluorescence
(FL) spectrum shows a good agreement in general. We therefore conclude
that our excitonic model captures correctly the features of the studied
system that are the most relevant in the present study. 

The FL spectrum in Fig. \ref{fig:Excitonic-model}B is dominated by
the lowest four excitons, which are the most populated ones. These
excitons are formed by strongly coupled pigments Chl 610-611-612,
Chl 602-603 and Chl 613-14 (see Ref. \cite{Liu2004} for nomenclature).
This is in agreement with previous modeling results for the trimeric
LHCII \cite{Novoderezhkin2005a}.

\begin{figure}[h]
\includegraphics{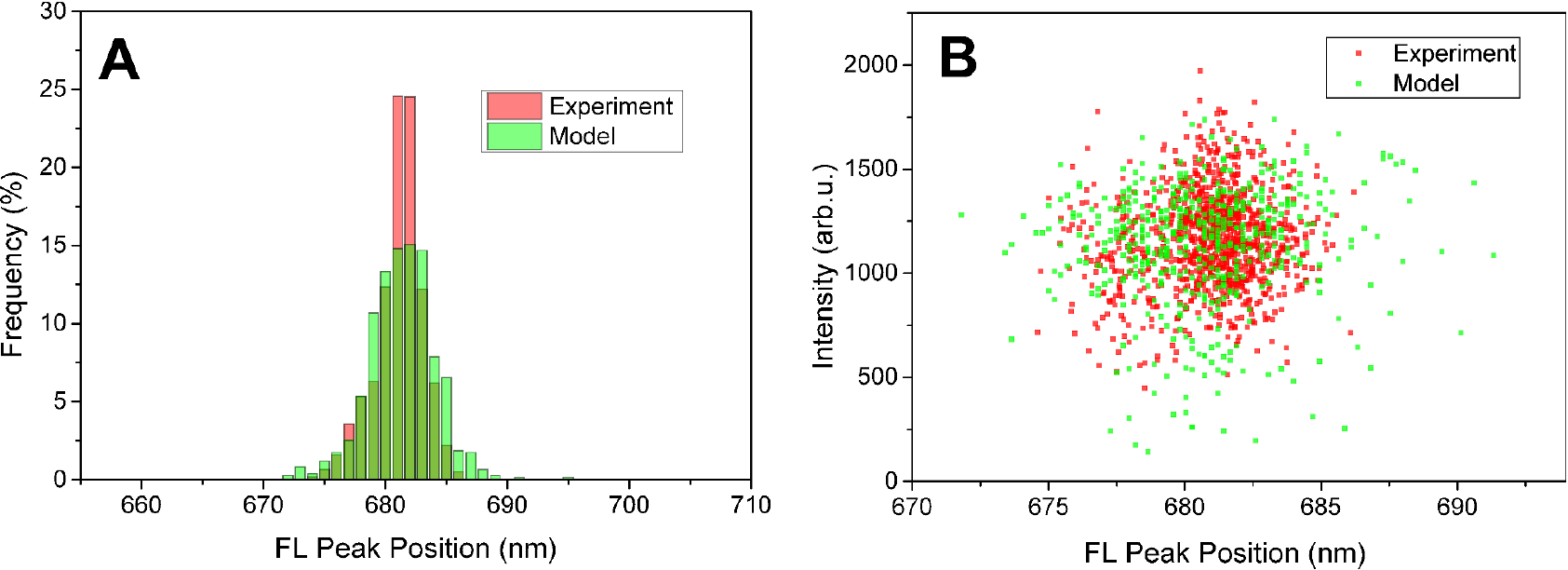}

\caption{\label{fig:FL-peak-distribution}Experimental (red) and calculated
(green) fluorescence peak distribution. (A) Peak position histogram,
(B) FL peak position and intensity plot. Theoretical points are calculated
as individual realizations of energetic disorder. }
\end{figure}

The calculated bulk spectra seem to be in a good agreement with the
experiment. Our model also reproduces successfully the statistics
from the SMS experiments. In Fig. \ref{fig:FL-peak-distribution}A
we present the FL spectral peak distribution compared to the experiment.
We can see that the calculated distribution is a little broader than
the experimental one, but the agreement is again reasonably good.
The larger distribution width of the calculated spectra can be explained
by a relatively long integration time in the experiment (1 s), during
which the system samples several individual realizations of the disorder.
In extreme cases the measured values get averaged towards the mean
value. As a result the measured distribution is narrower. From our
comparisons between experiments and calculations we conclude that
the exciton model with Gaussian disorder not only reproduces the averaged
absorption spectra and equilibriated populations of excitons (resulting
in characteristic FL spectra), but also the individual realizations
provided by this model are in a good agreement with the experiment.

The next feature of the LHCII single molecular spectra that we need
to address is the significant amount of blinking, i. e. reversible
switching to the off state. Since the measured dwell times in the
off state are in the range of seconds, which is significantly longer
than the lifetime of any long-living species such as triplet states
\cite{Schodel1998}, the off states must correspond to states with
efficient excitation energy dissipation. Correspondingly, our model
has to be extended by including some fluorescence quenching mechanism.

\subsubsection{Lutein Lut 1 as a Fluorescence Quencher\label{sub:Lutein-Lut-1}}

One of the possible mechanism of FL quenching in LHCII, proposed by
Ruban and coworkers \cite{Ruban2007,Duffy2013}, is an excitation
energy transfer from the lowest chl a exciton states to a lutein molecule,
Lut 1 (lut620), see Ref. \cite{Liu2004} for nomenclature). The Lut
1 molecule resides in the vicinity of the so-called terminal emitter
group of chlorophylls, composed of Chls a610, 611 and 612, and it
is supposed to be coupled mainly to Chl a612 \cite{Ruban2007,Duffy2013}.
The S1 state of the Lut molecule is optically forbidden, and it has
a short (10 ps) lifetime due to a decay through a non-radiative channel
\cite{Polivka2002}. The transition energy from the S0$\rightarrow$S1
of Lut is in the vicinity of the transition to the Chl $Q_{Y}$ state.
Due to its short excited state life-time, Lut could in principle act
as an excitation (and fluorescence) quencher. Let us test this mechanism
within our model to see if it can account for the observed blinking.
The important parameters of the lutein in context of our model are
its S1 state site energy relative to its groundstate and the coupling
to chlorophylls, in particular to Chl a612. 

\begin{figure}[h]
\includegraphics{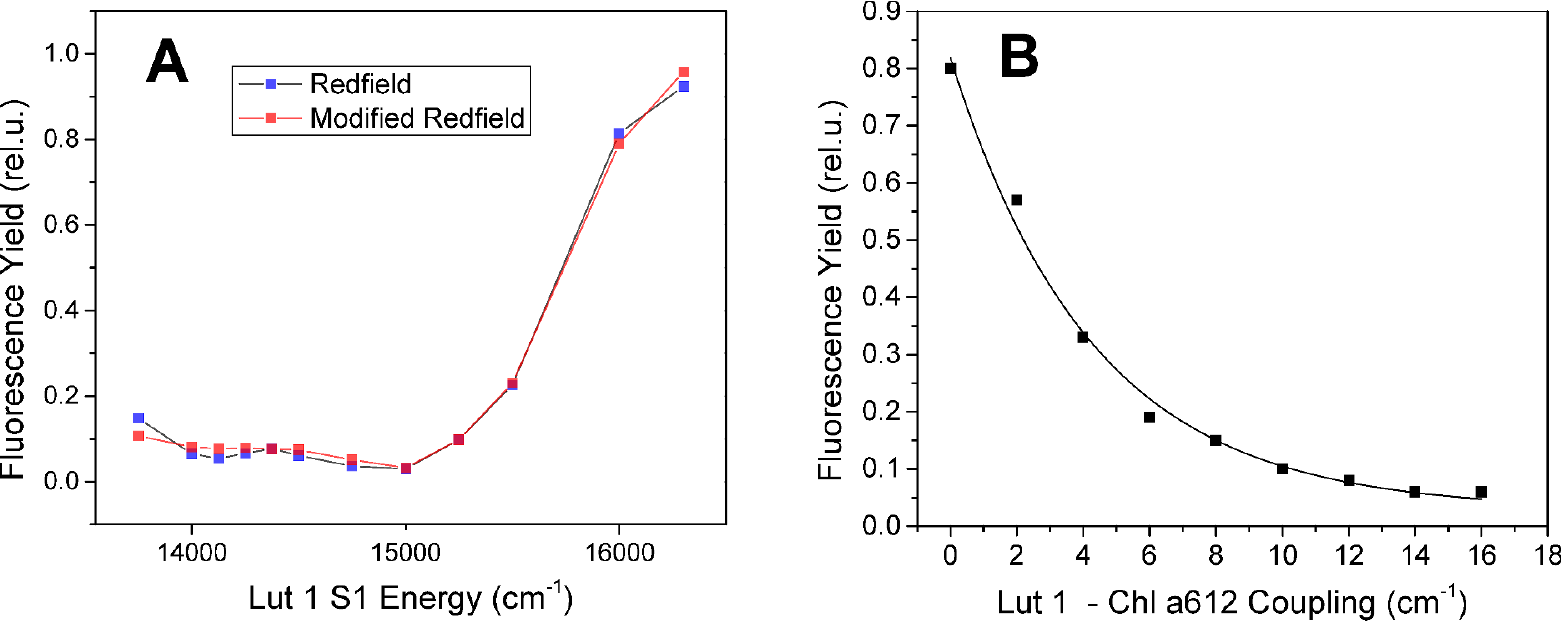}

\caption{\label{fig:The-Lut-1} Role of Lut 1 parameters. Dependence of the
fluorescence yield on (A) Lut 1 S1 energy and (B) Lut 1 - Chl a612
coupling. The energy dependence is calculated with Redfield (blue)
and Modified Redfield (red) theory for comparison. The dependence
on the coupling strength depicts calculated points (Redfield theory)
fitted with exponential dependence. Already a realistically small
coupling around $12\mbox{ cm}^{-1}$ leads to significant FL quenching.}
\end{figure}

In Fig. \ref{fig:The-Lut-1}A we present the dependence of the relative
FL quantum yield on the Lut energy for fixed value of the Lut-Chl
coupling of 14 cm$^{-1}$. The energy dependence agrees well with
the one obtained by Ruban \cite{Duffy2013}. The quenching is only
efficient when the Lut energy is below one of the red chlorophylls
(around 15100 cm$^{-1}$) and the plateau enables Lut to act as an
efficient quencher even in disordered systems. 

In Fig. \ref{fig:The-Lut-1}B we show the dependency of the FL quantum
yield on the Lut-Chl coupling for fixed Lut energy 14500 cm$^{-1}$.
Due to large reorganization energy, 14500 cm$^{-1}$ corresponds to
the zero-phonon line at 13900 cm$^{-1}$ and thus agrees with experimental
observations \cite{Polivka2002}. From the coupling dependence of
fluorescence in Fig. \ref{fig:The-Lut-1}B we can conclude that weak
coupling is sufficient for Lut 1 to act as a fluorescence quencher.
Very importantly, even small changes in the Lut-Chl coupling can result
in a big difference in the fluorescence intensity. Based on this analysis
we decided to use lutein S1 site energy of 14500 cm$^{-1}$ in our
simulations of blinking. We define the quenched state by the value
of 12 cm$^{-1}$ for coupling of the Lut 1 to Chl 612 and the unquenched
state by the zero coupling. Our model allows any type of time dependency
of the Chl-Lut coupling to be used, and it could in principle accommodate
input from structure based MD studies and quantum chemical treatment
of the (Dexter type) coupling of the Chl $Q_{Y}$ and the Lut S1 states.
However, a much rougher phenomenological model of the Chl-Lut coupling
changes enables a better discussion of the feasibility of the proposed
quenching mechanism than the parameter free \emph{ab initio} calculations
suffering from uncertainties in the structural information. Next we
proceed to model the blinking statistics.

\subsubsection{Model of the On-Off State Switching\label{sub:Model-On-Off}}

As mentioned in the Introduction, the blinking statistics alone can
be well described by a two-level model proposed by Valkunas et al.
in Ref. \cite{Valkunas2012}. By random fluctuations, the protein
samples its potential energy surface (PES) performing thus a random
walk (RW). The model of Ref. \cite{Valkunas2012} assumes that there
are two stable conformations of the protein corresponding to two minima
of the protein PES. These are approximated by two harmonic potentials.
The protein undergoes a RW in this potential, and at every step it
has a certain probability to switch from its current PES to the other
PES. In our treatment we use a discrete RW description, which enables
us to follow individual trajectories of the proteins. For the details
of the approach taken in this study and the differences from the original
model by Valkunas et al., see SI and Refs. \cite{Valkunas2012,Chmeliov2013}. 

\begin{figure}[h]
\includegraphics{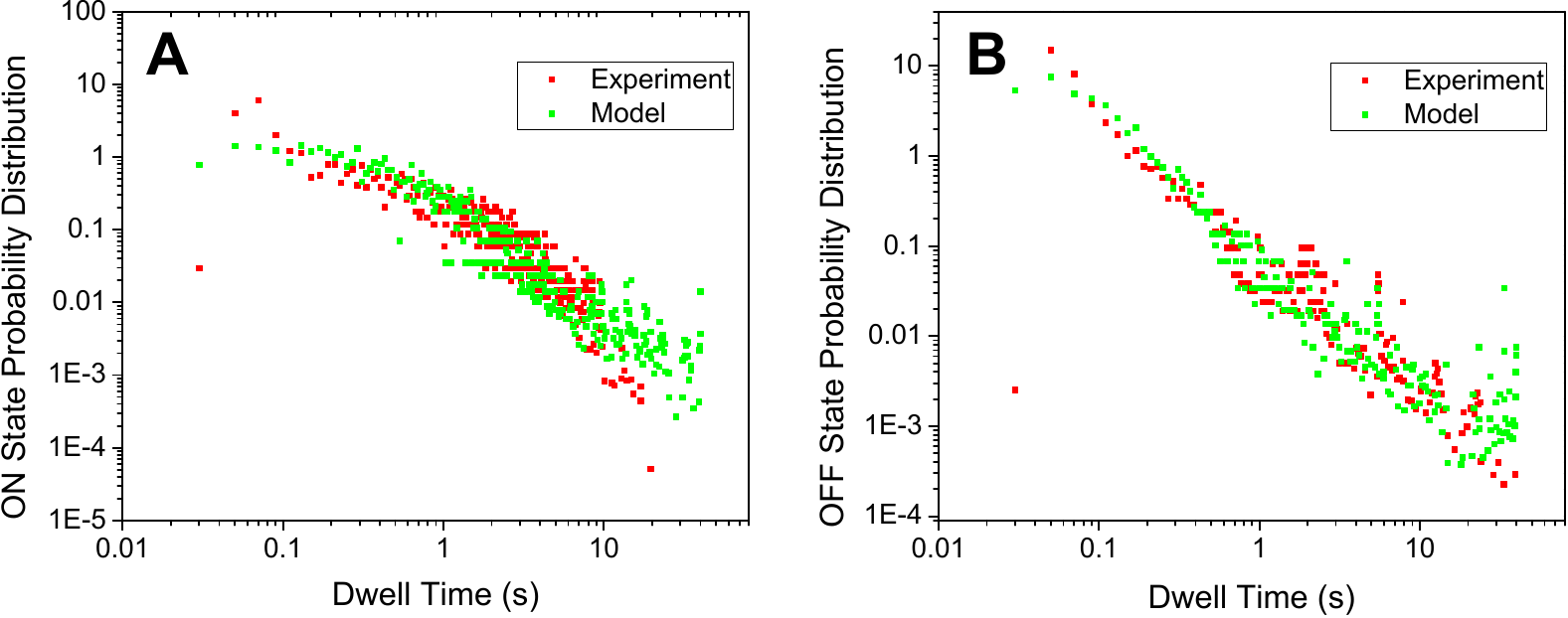}\caption{\label{fig:Dwell-times} Experimental (red) and calculated (green)
probability distribution of dwell times in the (A) ON and (B) OFF
state, logarithmic scale. While the OFF state distribution follows
a power law, the ON state distribution has an exponential tail at
longer times.}
\end{figure}

To connect the two PES model to our particular Lut quenching model,
we assume that the change of protein conformation somehow changes
the Lut-Chl coupling. The Lut S1 state does not have a dipole moment,
and the resonance coupling similar to those between allowed states
does not occur here. The two different protein conformations responsible
for the quenched/unquenched states would then result in two slightly
different orientations/positions of the pigments, leading to different
strengths of the coupling. This mechanism is in accord with recent
quantum-chemical study by Duffy et al. \cite{Chmeliov2015b}, where
small configurational changes were found to lead to substantial changes
in chl-car couplings. The switching between the PES is controlled
by the RW model with diffusion parameters adjusted to fit the experimental
dwell time distributions. The comparison between the calculated and
experimentally determined dwell-times is presented in Figs. \ref{fig:Dwell-times}A
and \ref{fig:Dwell-times}B, for the on and off times, respectively.
The agreement is again fairly good letting us believe that our phenomenological
model captures the most important features of the protein dynamics
affecting the blinking behaviour.

\subsubsection{Intensity traces\label{sub:Intensity-traces}}

\begin{figure}[h]
\includegraphics{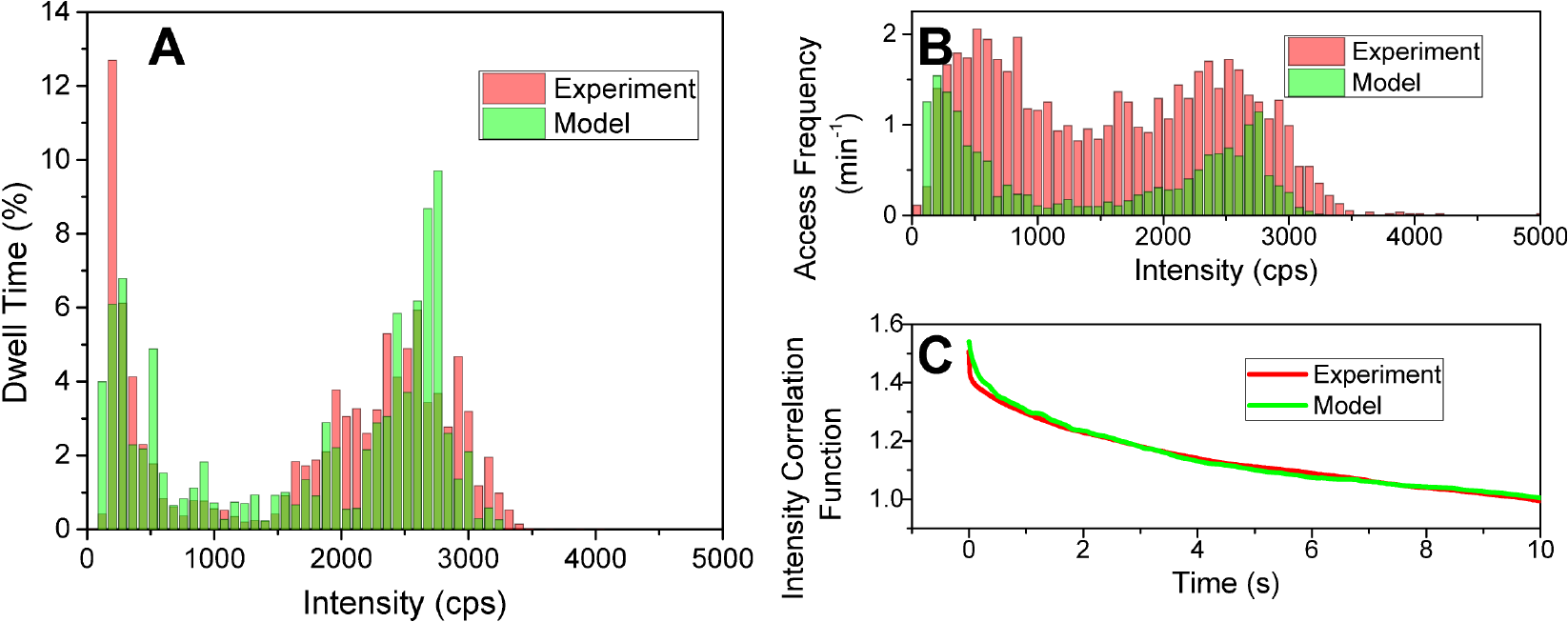}\caption{\label{fig:Intensity-traces-statistics} Experimental (red) and calcuated
(green) intensity blinking statistics. (A) The percentage of time
the complexes dwelled on respective intensity levels. The two-state
structure of the low-intensity OFF states and higher intensity ON
states is apparent. (B) How often per minute the complexes accessed
the respective intensity levels. The experimental frequency is higher
due to switching within the ON/OFF states. (C) The intensity correlation
function.}
\end{figure}

Finally, we can connect the two models described above and simulate
the blinking behaviour. To this end we continuously model the fluorescence
of the LHCII complexes, and output the intensity (and spectrum) every
10 ms corresponding to the experimental integration bins. Simultaneously,
we let the protein do the RW on its PES, and we adjust the Chl/Lut
coupling, when the protein switches between PES. To obtain more realistic
traces, either the site energies or couplings can be slightly varied
after every jump, resulting in different energy levels. Such a procedure,
however, does not lead to qualitatively different conclusions and
it can be in principle omitted. In the calculations presented here,
we used Gaussian disorder with the FWHM of 0.3 cm$^{-1}$ for couplings
and 1 cm$^{-1}$ for energies. For every realization of the disorder
a 60 s trace is modelled. This is repeated for 200 realizations of
the disorder, reflecting the experimental conditions. The resulting
statistics are presented in Fig. \ref{fig:Intensity-traces-statistics}. 

Total dwell times in Fig. \ref{fig:Intensity-traces-statistics}A
represent the overall amount of time spent in a given intensity level.
From this we can see the two-level character of the blinking and simultaneously
also the presence of some intermediate levels, which result from particular
disorder realizations. The dwell times are similar for the on and
off states. Fig. \ref{fig:Intensity-traces-statistics}B shows how
often are the intensity levels visited per fixed amount of time. The
modelled access frequency distribution is naturally very symmetrical,
a direct result of the fact that in the model the complex switches
only between the on and off state. The number of on/off states visited
per minute therefore has to be the same. The experimentally analyzed
intensity traces contain also jumps between levels within the on/off
states in an amount which can be, to some extent, modified by adjusting
the sensitivity of the level analysis. The presence of these intra-state
jumps results in higher switching frequency and wider distribution
in intensities, causing a moderate discrepancy between experiment
and simulation. In order to include this kind of switching into the
model, dynamic sampling of the disorder would have to be incorporated.
Work in this direction will be presented elsewhere. 

For the reasons stated above, the level access frequency distribution
is not well suited for comparison of the model with the experiment.
A more appropriate measure of the blinking would be the intensity-intensity
correlation function defined as $h^{(2)}(\tau)=\nicefrac{\left\langle I(\tau)I(0)\right\rangle }{\left\langle I(0)\right\rangle ^{2}}$.
This quantity is well-known from single-molecule measurements, where
it is often used to characterize the blinking behaviour \cite{SMSignalsBook}.
In Fig. \ref{fig:Intensity-traces-statistics}C we present $h^{(2)}(\tau)$
obtained from 50 measured long enough traces, compared with the model. We can
see that the agreement between experiment and theory is good, indicating
that our model gives reasonable switching between the intensity levels.
The shape of the correlation function is given by the dwell time statistics,
see also Fig. \ref{fig:Dwell-times}. The initial fast drop implies
the abundance of short blinking events. This results from the mechanism
of the protein switching between its potential surfaces, where the
short succesive blinking events are caused by the dynamics in the
vicinity of the PES intersection.

\section*{Discussion\label{sec:Discussion}}

In Section \ref{sub:Excited-State-Dynamics} of the Results we derived
approximate equations of motion for populations evolution on the timescale
of the SMS experiment. We have shown that the weak illumination regime,
in which the SMS experiment is performed, allows for an effective
source term description of the light-matter interaction in which short
time transient effects arising in the photoinduced evolution of molecular
systems can be consistently accommodated side by side with the slow
evolution of the protein bath observed in SMS experiments. We demonstrated
the validity and scope of application of our equations by simulating
our single molecule experiment on LHCII monomers. Based on the recent
research in elucidating the NPQ mechanism in LHCII and the connection
between fluorescence quenching and energy dissipation we implemented
energy transfer to Lut 1 as a blinking mechanism. Based on our calculations
we were able to confirm the findings of Duffy et al. \cite{Duffy2013}
and Chmeliov et al.\cite{Chmeliov2015b} that, within a reasonable
range of parameters, Lut 1 can indeed act as an efficient quencher.
Since our model extends the previous treatment by including realistic
excitation conditions and population transfer rates, it is remarkable,
how similar our fluorescence quenching dependence on the Lut 1 energy
is to the one in Ref. \cite{Duffy2013}. Moreover, we were able to
confirm that Lut 1 acts as an efficient quencher also under AM1.5
illumination (data not shown since the dependence is very similar).
At the same time we can see that the amount of quenching is very sensitive
to the change of coupling of the Lut to the chlorophylls. Since the
coupling itself is very sensitive to the distance and orientation
between the pigments, it provides a possible link to the protein conformation
change working as a switching mechanism as proposed in Ref. \cite{Valkunas2012,Chmeliov2015b}.
Indeed, when using the 2-level switching model to control the change
of coupling, we are able to reproduce the experimentally obtained
blinking statistics. Although far from being exclusive in any way,
our argument strongly supports the notion of the protein acting as
a conformational switch regulating the amount of quenching in the
system. 

The agreement between the theory and experiment also serves as a good
demonstration of the scope of our equations. They provide a description
for controlling the energy transfer in the system by modulating the
parameters of the excitonic model. We are aware of the remaining phenomenological
nature of our connection of the 2-level switching model with the excitonic
model. Further improvements in the direction of introducing more parameters
with a particular physical meaning, for example the relation to the
actual PES shape, are needed and will be subject to further study.
Also recent experimental observations indicate the presence of more
relevant timescales in the intensity traces suggesting the inadequacy
of a 2-level model with one reaction coordinate. Finally, although
some connection between the fluorescence blinking and NPQ was already
shown by Kr\"uger et al. \cite{Kruger2011}, their exact relation
is yet to be elucidated. The equations derived here are a suitable
tool for these future investigations.

\section*{}

\section*{Methods}

\subsection*{Sample preparation}

Trimeric LHCII complexes of spinach were isolated from freshly prepared
thylakoid membranes as described earlier \cite{VanRoon2000}. Monomeric
complexes were obtained by incubating LHCII trimers with 1\% (w/v)
octyl glucoside and 10 $\mathrm{\mu g/mL}$ phospholipase A2 (Sigma)
\cite{Nussberger1994}. Subsequent fast protein liquid chromatography
(FPLC) ensured a homogeneous sample preparation. The ensemble fluorescence
absorption and emission spectra were measured on a Lambda40 spectro-photometer
(Perkin-Elmer) and a FluoroLog Tau-3 (Jobin Yvon), respectively. For
SMS experiments, the sample was diluted down to a concentration of
\textasciitilde{} 10 pM in a measuring buffer (20 mM Hepes, pH 8 and
0.03\% (w/v) n-Dodecyl ${\textstyle \beta}$-D-maltoside) and then
immobilized on a PLL (poly-L-Lysine, Sigma) coated cover glass. Addition
of an oxygen scavenging mix of 750 $\mathrm{\mathrm{\mu g/ml}}$ Glucose
Oxidase, 100 $\mathrm{\mu g/ml}$ Catalase and 7.5 $\mathrm{mg/ml}$
Glucose (all Sigma) to the closed sample chamber inhibited the formation
of highly reactive singlet oxygen and improved the photostability
of complexes.

\subsection*{Single-molecule detection}

A confocal microscope was used to measure the fluorescence of single
complexes at 5 $^{\circ}C$ as described previously \cite{Kruger2010}.
The sample was excited at 630 nm utilizing a Ti:sapphire laser (Coherent
MIRA 900F) with a pulse width of 200 fs and a repetition rate of 76
MHz coupled to a tunable optical parametric oscillator (Coherent MIRA
OPO). Near-circular polarized light was obtained by utilizing a Berek
polarization compensator (5540 New Focus). A fluorescence beam splitter
(70:30, Thorlabs) allowed us to simultaneously measure the fluorescence
spectrum \textit{via} a CCD camera (Spec10:100BR, Roper Scientific)
with an integration time of one second and the wavelength integrated
fluorescence intensity \textit{via} an avalanche diode (SPCM-AQR 16,
Perkin Elmer) with a binning time of 10 milliseconds. The fluorescence
of one complex was analyzed for either one minute or until it photo-bleached
and a set of 200 complexes served as the basis for statistical analysis.
The fluorescence peak distribution was obtained by fitting of a skewed
Gaussian to the fluorescence spectrum as shown in Kruger et al. \cite{Kruger2010}
and the blinking analysis was performed equivalently to the algorithm
described elsewhere \cite{Kruger2011a}.

\subsection*{Dynamics simulation}

The equations of motion, Eq. (\ref{eq:PopEvolution}), have quasi
constant coefficients, and they can therefore be written in the form 

\begin{equation}
\frac{\partial\mathbf{P}}{\partial t}=\mathbf{{\cal M}}\mathbf{P}+\mathbf{S},\label{eq:matrix_eq-1}
\end{equation}
where ${\cal M}$ is a matrix of relaxation and population transfer
rates, and $\mathbf{S}$ are the source terms. Eq. (\ref{eq:matrix_eq-1})
can be solved analytically:

\begin{equation}
\mathbf{P}(t)=\mathbf{{\cal M}}^{-1}(e^{\mathbf{{\cal M}}(t-t_{0})}-1)\mathbf{S}+e^{\mathbf{{\cal M}}(t-t_{0})}\mathbf{P}(t_{0}).
\end{equation}

This expression enables us to find the populations at any time without
the need to solve for all the previous times. If we aim at steady
state, $t_{0}$ can be send to $-\infty$ and the populations then
depend only parametrically on time $t$. The weak dependence of $\bm{{\rm S}}$
and ${\cal M}$ on $t$ makes it possible to explain changes in the
populations of the emitting states of a molecular system due to slow
changes of the protein environment and the structure of the molecular
system.

\section*{Acknowledgments}

P.M., J.M.G. and R.v.G. were supported by the VU University and by
an Advanced Investigator grant from the European Research Council
(no. 267333, PHOTPROT) to R.v.G.; R.v.G. was also supported by the
Nederlandse Organisatie voor Wetenschappelijk Onderzoek, Council of
Chemical Sciences (NWO-CW) via a TOP-grant (700.58.305), and by the
EU FP7 project PAPETS (GA 323901). R.v.G. gratefully acknowledges
his Academy Professor grant from the Netherlands Royal Academy of
Sciences (KNAW). P.M. and T.M. received financial support from the
Czech Science Foundation (GACR), grant no. 14-25752S. T. M. acknowledges
the kind support of the Nederlandse Organisatie voor Wetenschappelijk
Onderzoek (NWO) visitor grant nr. 040.11.423 during his stay at the
VU Amsterdam. 

\setcounter{figure}{0}
\makeatletter  
\renewcommand{\thefigure}{S\@arabic\c@figure} 
\makeatother
\pagebreak

\section*{Supplementary Information}

\subsection*{Exciton model\label{sec:Exciton-model}}

The interaction of the pigments, i. e. the system, with the vibrations,
i. e. its environment, is treated by second order perturbation theory.
There the bath is completely described by its correlation function
or, equivalently, in the frequency domain, by the spectral density
of the bath vibrations. We use the spectral density from Ref. \cite{Novoderezhkin2004}.
It is constructed from one overdamped oscillator representing the
slow protein vibrations and 48 high-frequency intrapigment modes (for
more context for the spectral density and excitonic model see Ref.
\cite{MukamelBook}):

\begin{equation}
C''(\omega)=2\lambda_{0}\frac{\gamma_{0}\omega}{\omega^{2}+\gamma_{0}^{2}}+\sum_{i=1}^{48}2S_{j}\omega_{j}\frac{\omega_{j}^{2}\gamma_{j}\omega}{(\omega^{2}-\omega_{j}^{2})^{2}+\omega^{2}\gamma_{j}^{2}}.
\end{equation}
The (temperature dependent) correlation function given by this spectral
density is assumed to be uncorrelated between individual sites and
differs only between Chl a and Chl b, while the difference is in the
coupling strength $\nu_{n}=\nu_{a/b}$:

\begin{equation}
C_{n}(\omega)=\nu_{n}C(\omega)=\nu_{n}\left(1+\coth\left(\frac{\hbar\omega}{2k_{B}T}\right)\right)C''(\omega).\label{eq:Correl.Func.}
\end{equation}

\noindent The time-dependent correlation function is then obtained
by Fourier transform of (\ref{eq:Correl.Func.}). We use the same
values of the spectral density parameters as in Ref. \cite{Novoderezhkin2004}.
The difference between the vertical, Franck-Condon transition of the
pigments, which are called site energies in this text, and their 0-0
transitions is given by the reorganization energy due to the interaction
with the bath:

\begin{equation}
\lambda=\frac{1}{\pi}\int_{0}^{\infty}\frac{C''(\omega)}{\omega}.
\end{equation}

\noindent Because the pigments are strongly coupled, the preferred
basis of calculations is the excitonic basis in which the system Hamiltonian
is diagonal with eigenvalues, exciton energies, $\omega_{i0}$. All
quantities, including correlation functions $C(t)$, reorganization
energies $\lambda$ and transition dipole moments $\mu$ have to be
transformed into the excitonic basis:

\begin{align}
C_{i}(t) & =\sum_{n}|c_{i}^{n}|^{2}C_{n}(t),\\
\mu_{i0} & =\sum_{n}c_{i}^{n}\mu_{n0},
\end{align}

\begin{equation}
\lambda_{i}=\sum_{n}|c_{i}^{n}|^{4}\nu_{n}\lambda.
\end{equation}
The position of the zero phonon lines of the excitonic transitions
is then

\begin{equation}
\omega_{i0}^{ZPL}=\omega_{i0}-\lambda_{i}.
\end{equation}

\noindent The spectral lineshapes are calculated by 2nd order cumulant
expansion employing the so-called lineshape functions: 

\begin{equation}
g_{ii}(t)=\int_{0}^{t}d\tau\int_{0}^{\tau}d\tau'C_{i}(\tau').
\end{equation}
The lineshape function is conveniently expressed in terms of the spectral
densities

\begin{equation}
g_{ii}(t)=\frac{1}{\pi}\int_{0}^{\infty}d\omega\frac{C_{i}''(\omega)}{\omega^{2}}\left[\coth\left(\frac{\hbar\omega}{2k_{B}T}\right)\left(1-\cos(\omega t)\right)+i\left(\sin(\omega t)-\omega t\right)\right].
\end{equation}

\noindent The absorption spectrum is calculated as

\begin{equation}
abs(\omega)\propto\omega\sum_{i}\chi_{i}(\omega),
\end{equation}
where the absorption lineshape is 

\begin{equation}
\chi_{i}(\omega)=|\mu_{i0}|^{2}{\rm Re}\int_{0}^{\infty}d\tau\:e^{-i(\omega-\omega_{i0})\tau-g_{ii}(\tau)-\frac{\Gamma_{i}}{2}\tau}.
\end{equation}

\noindent Here $\Gamma_{i}$ is the population relaxation rate from
state $i$. The fluorescence is similarly given as

\begin{equation}
FL(\omega)\propto\omega^{3}\sum_{i}P_{i}\tilde{\chi}_{i}(\omega),
\end{equation}
where $P_{i}$ is the steady-state population of state $i$ and the
fluorescence lineshape is 

\begin{equation}
\tilde{\chi}_{i}(\omega)=|\mu_{i0}|^{2}{\rm Re}\int_{0}^{\infty}d\tau\:e^{-i(\omega-\omega_{i0}+2\lambda_{i})\tau-g_{ii}^{*}(\tau)-\frac{\Gamma_{i}}{2}\tau}.
\end{equation}

\noindent The population transfer rates in the Redfield theory are
obtained as 

\begin{equation}
k_{ij}=\sum_{n}|c_{i}^{n}|^{2}|c_{j}^{n}|^{2}C_{n}(\omega_{i0}^{ZPL}-\omega_{j0}^{ZPL}).
\end{equation}

\noindent The population relaxation rates of chlorophylls are a result
of energy transfer and radiative and non-radiative decay, i.e. $\Gamma_{i}=\sum_{j}k_{ji}+\tilde{\Gamma}_{i}$,
where $\tilde{\Gamma}_{i}=\sum_{n}|c_{i}^{n}|^{2}\tilde{\Gamma}_{n}$,
$\tilde{\Gamma}_{n}=1/\tau_{n}$ is inverse lifetime $\tau_{n}$ of
site $n$ excited state, which is taken to be 3 ns for all $n$.

\subsection*{Bulk spectra measurement}

\begin{figure}[h]
\begin{centering}
\includegraphics{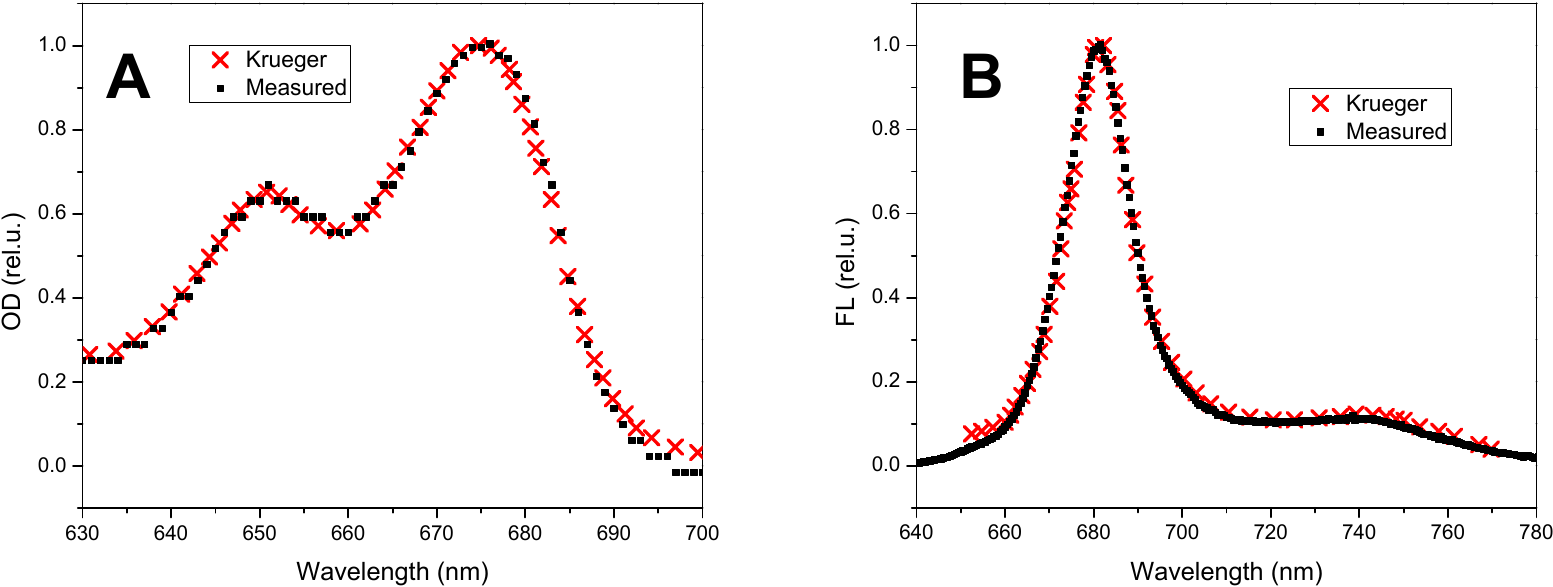}
\par\end{centering}

\caption{Bulk spectra measured and taken from Ref. \cite{Kruger2010}\label{fig:Bulk-spectra-measured}}
\end{figure}

To check for sample degradation a control bulk measurement of absorption
and fluorescence was performed. In Fig. \ref{fig:Bulk-spectra-measured}
the measured spectra are given together with the spectra of LHCII
trimers taken from \cite{Kruger2010}. Their excellent agreement confirms
absence of degradation in course of the experiment and also justifies
the usage of the trimer spectra for the bulk spectra modelling in
the main text. 

\subsection*{Random Walk Model\label{sec:RW-model}}

In the original description by Valkunas \cite{Valkunas2012} and,
in more detail, in \cite{Chmeliov2013}, the protein diffusive motion
was described as a continuous-time random walk (CTRW) on a two-dimensional
potential energy surface. Here we somewhat simplify this description
in the following way. The two generalized coordinates represent fast
and slow degrees of freedom and thus, when following the slow dynamics,
the fast fluctuations can be adiabatically eliminated. It can be shown
that when we consider the potential dependence on the fast coordinate
the same in both the on and off states for simplicity, i.e. setting
$x_{0}=0,\:\nicefrac{\lambda_{2}}{\lambda_{1}}=1$ in \cite{Valkunas2012},
the fast coordinate can be completely removed. This leaves us with
effectively one-dimensional problem. 

After transforming into dimensionless coordinates $y\sqrt{\frac{\gamma_{1}}{k_{B}T}}\rightarrow y$,
the parabolically approximated minima of the potential for the on
(1) and off (2) state are 

\begin{eqnarray}
\frac{U_{1}}{k_{B}T} & = & \frac{1}{2}y^{2},\nonumber \\
\frac{U_{2}}{k_{B}T} & = & \frac{1}{2}\frac{\gamma_{2}}{\gamma_{1}}\left(y-y_{0}\right)^{2}+\frac{U_{0}}{k_{B}T}.\label{eq:ypotentials-1}
\end{eqnarray}

\noindent Here $\gamma_{1,2}$ determine the steepness of the potential
and the second minimum is shifted by $y_{0}$ along the slow coordinate
and by $U_{0}$ along the potential energy.

The protein then performs random walk on the respective surface, with
probabilities to tunnel to the other surface

\begin{eqnarray}
\kappa_{1\rightarrow2}(y) & =k_{1} & e^{-\alpha|U_{1}-U_{2}|/\hbar\omega_{0}}\mbox{min\ensuremath{\{1,e^{(U_{1}-U_{2})/k_{B}T}}}\},\nonumber \\
\kappa_{2\rightarrow1}(y) & =k_{2} & e^{-\alpha|U_{1}-U_{2}|/\hbar\omega_{0}}\mbox{min\ensuremath{\{1,e^{(U_{2}-U_{1})/k_{B}T}}}\}.\label{eq:jumprob-1}
\end{eqnarray}

\noindent $k_{i}$ is a rate of falling from the $i-th$ potential,
the first exponential term reflects the energy gap law and the min
term ensures detailed balance condition. The coefficient $\alpha$
can be treated as a constant and $\omega_{0}$ is a characteristic
frequency of the protein environment vibrations responsible for the
tunneling.

The protein diffusion under this conditions can be described either
by the CTRW or by a discrete random walk (RW). The former approach
was employed by Valkunas et al.\cite{Valkunas2012,Chmeliov2013}.
However, we believe that for our purpose it is better to solve this
problem as a discrete RW for two reasons. First, the coupled Smoluchovski
equations for the CTRW on the two potentials can be decoupled only
for conditional probabilities, i.e. assuming that the system was in
the opposite state in the previous interval, and, in the same time,
employing the same, equilibrium initial condition for each dwell time.
When continuously modelling the trajectory of a single protein, we
do not have to include the resetting after switching and also the
conditioning will be inherent, as the system is observed being in
a particular state. And second, when we want to simulate the individual
intensity time traces, it is more natural to really follow the trajectories
of the individual proteins on their PES.

If we want to follow the time trace of every single molecule, we should
follow its particular trajectory. The blinking statistics will then
be recovered by averaging over a large number of molecules, exactly
as in the experiment. To the purpose of following trajectories of
individual proteins, we need to describe its discrete RW (DRW) in
the potential. We will denote probability of going right (left) as
$p$ ($q$). In the symmetrical RW we have $p=q=\frac{1}{2}$. If
the protein is at a coordinate $y$, in the next step it will move
with probability $p$ to $y+a$ and with probability $q$ to $y-a$,
where $a$ is the length of the step. Inspired by classical approach
by van Kampen \cite{vanKampen}, we augment the position dependent
probabilities in the presence of the potential $U(y)$ as 

\begin{eqnarray}
p(y) & = & \frac{1}{2}e^{-\frac{1}{k_{B}T}\left(U(y+a)-U(y)\right)},\nonumber \\
q(y) & = & \frac{1}{2}e^{-\frac{1}{k_{B}T}\left(U(y-a)-U(y)\right)}.\label{eq:pprobandqprob}
\end{eqnarray}

\noindent We note that defined probabilities defined in Eq. (\ref{eq:pprobandqprob})
reflect the detailed balance condition $\frac{p(y)}{q(y)}=e^{-\frac{1}{k_{B}T}\left(U(y+a)-U(y-a)\right)}$.
Considering a small step $a$, we can use Taylor expansion in $y$
$\left(e^{-\frac{1}{k_{B}T}U(y\pm a)}\approx e^{-\frac{1}{k_{B}T}U(y)}\left(1\mp\frac{1}{k_{B}T}\frac{dU}{dy}(y)a\right)\right)$,
obtaining

\begin{equation}
p(y)-q(y)=-\frac{1}{k_{B}T}\frac{dU}{dy}(y)a.
\end{equation}

\noindent Now, considering that $p(y)+q(y)=1$, we get 

\begin{equation}
p(y)=\frac{1}{2}-\frac{1}{2k_{B}T}\frac{dU(y)}{dy}a.
\end{equation}

\noindent Using the potential form (\ref{eq:ypotentials-1}), we get
for the probabilities 

\begin{eqnarray}
p_{1}(y) & = & \frac{1}{2}-\frac{1}{2k_{B}T}ya_{1},\nonumber \\
p_{2}(y) & = & \frac{1}{2}-\frac{1}{2k_{B}T}\frac{\gamma_{2}}{\gamma_{1}}(y-y_{0})a_{2}.\label{eq:probgo}
\end{eqnarray}

\noindent The length of the step $a$ on respective surface can be
related to the transformed diffusion coefficient $D_{1,2}$: 
\begin{equation}
a_{1,2}=\sqrt{\frac{D_{1,2}\gamma_{1}}{k_{B}T}\Delta t},
\end{equation}

\noindent where $\Delta t$ is the time duration of the step. During
$\Delta t$ the protein walks either to the left or right with the
respective probability and with the probability $\kappa_{i\rightarrow j}(y)\Delta t$
switches to the other surface.

Similarly to \cite{Valkunas2012} the position near the potential
intersection can be chosen as an initial condition. However, as we
do not include resetting after switching in our approach, this determines
only the starting point of each trajectory and is therefore not of
significant importance. In Table \ref{tab:RW-model-paramaters} we
present the parameters of  our RW model compared to the ones used
by Valkunas et al \cite{Valkunas2012}. Considering the differences
- LHCII monomers vs trimers, discrete vs continuous RW, no resetting
vs resetting - the agreement is satisfactory.

\begin{table}[h]
\begin{centering}
\begin{tabular}{|c|c|c|}
\hline 
parameter & Valkunas \cite[pH 8]{Valkunas2012} & This work\tabularnewline
\hline 
\hline 
$\nicefrac{\lambda_{2}}{\lambda_{1}}$ & 0.2 & 1.0\tabularnewline
\hline 
$\nicefrac{\gamma_{2}}{\gamma_{1}}$ & 0.72 & 0.72\tabularnewline
\hline 
$x_{0}\sqrt{\nicefrac{\lambda_{1}}{k_{B}T}}$ & 1.0 & 0.0\tabularnewline
\hline 
$y_{0}\sqrt{\nicefrac{\gamma_{1}}{k_{B}T}}$ & 8.57 & 6.07\tabularnewline
\hline 
$\nicefrac{U_{0}}{k_{B}T}$ & 1.5 & 1.5\tabularnewline
\hline 
$k_{1}^{-1}$ & 430 ms & 330 ms\tabularnewline
\hline 
$k_{2}^{-1}$ & 4.8 ms & 50 ms\tabularnewline
\hline 
$\left(D_{1}\nicefrac{\gamma_{1}}{k_{B}T}\right)^{-1}$ & 3.8 s & 66 s\tabularnewline
\hline 
$\left(D_{2}\nicefrac{\gamma_{1}}{k_{B}T}\right)^{-1}$ & 1.4 s & 10 s\tabularnewline
\hline 
$\nicefrac{\hbar\omega_{0}}{\alpha k_{B}T}$ & 1.0 & 1.0\tabularnewline
\hline 
\end{tabular}
\par\end{centering}

\caption{RW model parameters\label{tab:RW-model-paramaters}}
\end{table}


\begin{thebibliography}{10}
\expandafter\ifx\csname url\endcsname\relax
  \def\url#1{\texttt{#1}}\fi
\expandafter\ifx\csname urlprefix\endcsname\relax\def\urlprefix{URL }\fi
\providecommand{\bibinfo}[2]{#2}
\providecommand{\eprint}[2][]{\url{#2}}

\bibitem{BlankenshipBook}
\bibinfo{author}{Blankenship, R.~E.}
\newblock \emph{\bibinfo{title}{{Molecular Mechanisms of Photosynthesis}}}
  (\bibinfo{publisher}{Blackwell Science}, \bibinfo{year}{2002}).

\bibitem{Wraight1974}
\bibinfo{author}{Wraight, C.~A.} \& \bibinfo{author}{Clayton, R.~K.}
\newblock \bibinfo{title}{{The absolute quantum efficiency of
  bacteriochlorophyll photooxidation in reaction centres of Rhodopseudomonas
  spheroides}}.
\newblock \emph{\bibinfo{journal}{Biochim. Biophys. Acta - Bioenerg.}}
  \textbf{\bibinfo{volume}{333}}, \bibinfo{pages}{246--260}
  (\bibinfo{year}{1974}).

\bibitem{Blankenship2011}
\bibinfo{author}{Blankenship, R.~E.} \emph{et~al.}
\newblock \bibinfo{title}{{Comparing photosynthetic and photovoltaic
  efficiencies and recognizing the potential for improvement.}}
\newblock \emph{\bibinfo{journal}{Science}} \textbf{\bibinfo{volume}{332}},
  \bibinfo{pages}{805--809} (\bibinfo{year}{2011}).

\bibitem{VanAmerongen2000}
\bibinfo{author}{van Amerongen, H.}, \bibinfo{author}{van Grondelle, R.} \&
  \bibinfo{author}{Valkunas, L.}
\newblock \emph{\bibinfo{title}{{Photosynthetic Excitons}}}
  (\bibinfo{publisher}{World Scientific}, \bibinfo{address}{London},
  \bibinfo{year}{2000}).

\bibitem{Ruban2012}
\bibinfo{author}{Ruban, A.~V.}, \bibinfo{author}{Johnson, M.~P.} \&
  \bibinfo{author}{Duffy, C. D.~P.}
\newblock \bibinfo{title}{{The photoprotective molecular switch in the
  photosystem II antenna}}.
\newblock \emph{\bibinfo{journal}{Biochim. Biophys. Acta - Bioenerg.}}
  \textbf{\bibinfo{volume}{1817}}, \bibinfo{pages}{167--181}
  (\bibinfo{year}{2012}).

\bibitem{Holt2005}
\bibinfo{author}{Holt, N.~E.} \emph{et~al.}
\newblock \bibinfo{title}{{Carotenoid cation formation and the regulation of
  photosynthetic light harvesting.}}
\newblock \emph{\bibinfo{journal}{Science}} \textbf{\bibinfo{volume}{307}},
  \bibinfo{pages}{433--436} (\bibinfo{year}{2005}).

\bibitem{Ruban2007}
\bibinfo{author}{Ruban, A.~V.} \emph{et~al.}
\newblock \bibinfo{title}{{Identification of a mechanism of photoprotective
  energy dissipation in higher plants.}}
\newblock \emph{\bibinfo{journal}{Nature}} \textbf{\bibinfo{volume}{450}},
  \bibinfo{pages}{575--8} (\bibinfo{year}{2007}).

\bibitem{Avenson2008}
\bibinfo{author}{Avenson, T.~J.} \emph{et~al.}
\newblock \bibinfo{title}{{Zeaxanthin radical cation formation in minor
  light-harvesting complexes of higher plant antenna}}.
\newblock \emph{\bibinfo{journal}{J. Biol. Chem.}}
  \textbf{\bibinfo{volume}{283}}, \bibinfo{pages}{3550--3558}
  (\bibinfo{year}{2008}).

\bibitem{Bode2009}
\bibinfo{author}{Bode, S.} \emph{et~al.}
\newblock \bibinfo{title}{{On the regulation of photosynthesis by excitonic
  interactions between carotenoids and chlorophylls.}}
\newblock \emph{\bibinfo{journal}{Proc. Natl. Acad. Sci. U. S. A.}}
  \textbf{\bibinfo{volume}{106}}, \bibinfo{pages}{12311--12316}
  (\bibinfo{year}{2009}).

\bibitem{Berera2010}
\bibinfo{author}{Berera, R.}, \bibinfo{author}{{Van Stokkum}, I. H.~M.},
  \bibinfo{author}{Kennis, J. T.~M.}, \bibinfo{author}{Grondelle, R.~V.} \&
  \bibinfo{author}{Dekker, J.~P.}
\newblock \bibinfo{title}{{The light-harvesting function of carotenoids in the
  cyanobacterial stress-inducible IsiA complex}}.
\newblock \emph{\bibinfo{journal}{Chem. Phys.}} \textbf{\bibinfo{volume}{373}},
  \bibinfo{pages}{65--70} (\bibinfo{year}{2010}).

\bibitem{Muller2010}
\bibinfo{author}{M{\"{u}}ller, M.~G.} \emph{et~al.}
\newblock \bibinfo{title}{{Singlet energy dissipation in the photosystem II
  light-harvesting complex does not involve energy transfer to carotenoids.}}
\newblock \emph{\bibinfo{journal}{Chemphyschem}} \textbf{\bibinfo{volume}{11}},
  \bibinfo{pages}{1289--96} (\bibinfo{year}{2010}).

\bibitem{Fuciman2012}
\bibinfo{author}{Fuciman, M.} \emph{et~al.}
\newblock \bibinfo{title}{{The Role of Xanthophylls in Light-Harvesting in
  Green Plants: A Spectroscopic Investigation of Mutant LHCII and Lhcb
  Pigment-Proteins Complexes.}}
\newblock \emph{\bibinfo{journal}{J. Phys. Chem. B}}
  \textbf{\bibinfo{volume}{116}}, \bibinfo{pages}{3834--49}
  (\bibinfo{year}{2012}).

\bibitem{Staleva2015}
\bibinfo{author}{Staleva, H.} \emph{et~al.}
\newblock \bibinfo{title}{{Mechanism of photoprotection in the cyanobacterial
  ancestor of plant antenna proteins}}.
\newblock \emph{\bibinfo{journal}{Nat. Chem. Biol.}}
  \textbf{\bibinfo{volume}{11}} (\bibinfo{year}{2015}).

\bibitem{Haarer1990}
\bibinfo{author}{Haarer, D.} \& \bibinfo{author}{Silbey, R.}
\newblock \bibinfo{title}{{Hole-Burning Spectroscopy of Glasses}}.
\newblock \emph{\bibinfo{journal}{Phys. Today}} \textbf{\bibinfo{volume}{43}},
  \bibinfo{pages}{58} (\bibinfo{year}{1990}).

\bibitem{Johnson1989}
\bibinfo{author}{Johnson, S.}, \bibinfo{author}{Tang, D.},
  \bibinfo{author}{Jankowiak, R.}, \bibinfo{author}{Hayes, J.~M.} \&
  \bibinfo{author}{Small, G.~J.}
\newblock \bibinfo{title}{{Structure and marker mode of the primary electron
  donor state absorption of photosynthetic bacteria: hole-burned spectra}}.
\newblock \emph{\bibinfo{journal}{J. Phys. Chem.}}
  \textbf{\bibinfo{volume}{93}}, \bibinfo{pages}{5953--5957}
  (\bibinfo{year}{1989}).

\bibitem{Creemers1999}
\bibinfo{author}{Creemers, T. M.~H.}, \bibinfo{author}{Caro, C. A.~D.},
  \bibinfo{author}{Visschers, R.~W.}, \bibinfo{author}{Grondelle, R.~V.} \&
  \bibinfo{author}{Vo, S.}
\newblock \bibinfo{title}{{Spectral Hole Burning and Fluorescence Line
  Narrowing in Subunits of the Light-Harvesting Complex LH1 of Purple
  Bacteria}}.
\newblock \emph{\bibinfo{journal}{J. Phys. Chem. B}}
  \textbf{\bibinfo{volume}{103}}, \bibinfo{pages}{9770--9776}
  (\bibinfo{year}{1999}).

\bibitem{Jonas2003}
\bibinfo{author}{Jonas, D.~M.}
\newblock \bibinfo{title}{{Two-dimensional femtosecond spectroscopy.}}
\newblock \emph{\bibinfo{journal}{Annu. Rev. Phys. Chem.}}
  \textbf{\bibinfo{volume}{54}}, \bibinfo{pages}{425--463}
  (\bibinfo{year}{2003}).

\bibitem{Moerner1989}
\bibinfo{author}{Moerner, W.} \& \bibinfo{author}{Kador, L.}
\newblock \bibinfo{title}{{Optical detection and spectroscopy of single
  molecules in a solid}}.
\newblock \emph{\bibinfo{journal}{Phys. Rev. Lett.}}
  \textbf{\bibinfo{volume}{62}}, \bibinfo{pages}{2535--2538}
  (\bibinfo{year}{1989}).

\bibitem{Schlau-Cohen2015}
\bibinfo{author}{Schlau-Cohen, G.~S.} \emph{et~al.}
\newblock \bibinfo{title}{{Single-Molecule Identification of Quenched and
  Unquenched States of LHCII}}.
\newblock \emph{\bibinfo{journal}{J. Phys. Chem. Lett.}}
  \bibinfo{pages}{860--867} (\bibinfo{year}{2015}).

\bibitem{Liu2004}
\bibinfo{author}{Liu, Z.} \emph{et~al.}
\newblock \bibinfo{title}{{Crystal structure of spinach major light-harvesting
  complex at 2 . 72 A resolution}}.
\newblock \emph{\bibinfo{journal}{Nature}} \textbf{\bibinfo{volume}{428}},
  \bibinfo{pages}{287--292} (\bibinfo{year}{2004}).

\bibitem{Novoderezhkin2004}
\bibinfo{author}{Novoderezhkin, V.~I.}, \bibinfo{author}{Palacios, M.~A.},
  \bibinfo{author}{van Amerongen, H.} \& \bibinfo{author}{van Grondelle, R.}
\newblock \bibinfo{title}{{Energy-Transfer Dynamics in the LHCII Complex of
  Higher Plants: Modified Redfield Approach}}.
\newblock \emph{\bibinfo{journal}{J. Phys. Chem. B}}
  \textbf{\bibinfo{volume}{108}}, \bibinfo{pages}{10363}
  (\bibinfo{year}{2004}).

\bibitem{Novoderezhkin2005a}
\bibinfo{author}{Novoderezhkin, V.~I.}, \bibinfo{author}{Palacios, M.~A.},
  \bibinfo{author}{van Amerongen, H.} \& \bibinfo{author}{van Grondelle, R.}
\newblock \bibinfo{title}{{Excitation dynamics in the LHCII complex of higher
  plants: modeling based on the 2.72 Angstrom crystal structure.}}
\newblock \emph{\bibinfo{journal}{J. Phys. Chem. B}}
  \textbf{\bibinfo{volume}{109}}, \bibinfo{pages}{10493--504}
  (\bibinfo{year}{2005}).

\bibitem{Novoderezhkin2011}
\bibinfo{author}{Novoderezhkin, V.}, \bibinfo{author}{Marin, A.} \&
  \bibinfo{author}{Grondelle, R.~V.}
\newblock \bibinfo{title}{{Supplementary Information Intra- and inter-monomeric
  transfers in the light harvesting LHCII complex : the Redfield-F{\"{o}}rster
  picture}}  (\bibinfo{year}{2011}).

\bibitem{Kruger2010}
\bibinfo{author}{Kr{\"{u}}ger, T. P.~J.}, \bibinfo{author}{Novoderezhkin,
  V.~I.}, \bibinfo{author}{Ilioaia, C.} \& \bibinfo{author}{van Grondelle, R.}
\newblock \bibinfo{title}{{Fluorescence spectral dynamics of single LHCII
  trimers.}}
\newblock \emph{\bibinfo{journal}{Biophys. J.}} \textbf{\bibinfo{volume}{98}},
  \bibinfo{pages}{3093--101} (\bibinfo{year}{2010}).

\bibitem{Tietz2001}
\bibinfo{author}{Tietz, C.} \emph{et~al.}
\newblock \bibinfo{title}{{Single molecule spectroscopy on the light-harvesting
  complex II of higher plants.}}
\newblock \emph{\bibinfo{journal}{Biophys. J.}} \textbf{\bibinfo{volume}{81}},
  \bibinfo{pages}{556--62} (\bibinfo{year}{2001}).

\bibitem{Kruger2011}
\bibinfo{author}{Kr{\"{u}}ger, T. P.~J.}, \bibinfo{author}{Ilioaia, C.},
  \bibinfo{author}{Valkunas, L.} \& \bibinfo{author}{van Grondelle, R.}
\newblock \bibinfo{title}{{Fluorescence intermittency from the main plant
  light-harvesting complex: sensitivity to the local environment.}}
\newblock \emph{\bibinfo{journal}{J. Phys. Chem. B}}
  \textbf{\bibinfo{volume}{115}}, \bibinfo{pages}{5083--95}
  (\bibinfo{year}{2011}).

\bibitem{Kruger2012}
\bibinfo{author}{Kr{\"{u}}ger, T. P.~J.} \emph{et~al.}
\newblock \bibinfo{title}{{Controlled disorder in plant light-harvesting
  complex II explains its photoprotective role.}}
\newblock \emph{\bibinfo{journal}{Biophys. J.}} \textbf{\bibinfo{volume}{102}},
  \bibinfo{pages}{2669--76} (\bibinfo{year}{2012}).

\bibitem{Kruger2011a}
\bibinfo{author}{Kr{\"{u}}ger, T. P.~J.}, \bibinfo{author}{Ilioaia, C.} \&
  \bibinfo{author}{van Grondelle, R.}
\newblock \bibinfo{title}{{Fluorescence intermittency from the main plant
  light-harvesting complex: resolving shifts between intensity levels.}}
\newblock \emph{\bibinfo{journal}{J. Phys. Chem. B}}
  \textbf{\bibinfo{volume}{115}}, \bibinfo{pages}{5071--82}
  (\bibinfo{year}{2011}).

\bibitem{Maly2015}
\bibinfo{author}{Mal{\'{y}}, P.}, \bibinfo{author}{Gruber, J.~M.},
  \bibinfo{author}{Cogdell, R.~J.}, \bibinfo{author}{Man{\v{c}}al, T.} \&
  \bibinfo{author}{van Grondelle, R.}
\newblock \bibinfo{title}{{Ultrafast Energy Relaxation in Single
  Light-Harvesting Complexes}}.
\newblock \emph{\bibinfo{journal}{arXiv:1511.04936[physics.bio-ph]}}
  (\bibinfo{year}{2015}).

\bibitem{MayKuhn}
\bibinfo{author}{May, V.} \& \bibinfo{author}{K{\"{u}}hn, O.}
\newblock \emph{\bibinfo{title}{{Charge and Energy Transfer Dynamics in
  Molecular Systems}}} (\bibinfo{publisher}{Wiley-VCH}, \bibinfo{year}{2011}).

\bibitem{MancalBook}
\bibinfo{author}{Valkunas, L.}, \bibinfo{author}{Abramavicius, D.} \&
  \bibinfo{author}{Man{\v{c}}al, T.}
\newblock \emph{\bibinfo{title}{{Molecular Excitation Dynamics and
  Relaxation}}} (\bibinfo{publisher}{Wiley-VCH}, \bibinfo{year}{2013}).

\bibitem{Ishizaki2009}
\bibinfo{author}{Ishizaki, A.} \& \bibinfo{author}{Fleming, G.~R.}
\newblock \bibinfo{title}{{Unified treatment of quantum coherent and incoherent
  hopping dynamics in electronic energy transfer: reduced hierarchy equation
  approach.}}
\newblock \emph{\bibinfo{journal}{J. Chem. Phys.}}
  \textbf{\bibinfo{volume}{130}}, \bibinfo{pages}{234111}
  (\bibinfo{year}{2009}).

\bibitem{Hein2012}
\bibinfo{author}{Hein, B.}, \bibinfo{author}{Kreisbeck, C.},
  \bibinfo{author}{Kramer, T.} \& \bibinfo{author}{Rodr{\'{\i}}guez, M.}
\newblock \bibinfo{title}{{Modelling of oscillations in two-dimensional
  echo-spectra of the Fenna-Matthews-Olson complex}}.
\newblock \emph{\bibinfo{journal}{New J. Phys.}} \textbf{\bibinfo{volume}{14}},
  \bibinfo{pages}{023018} (\bibinfo{year}{2012}).

\bibitem{Wilkins2015}
\bibinfo{author}{Wilkins, D.~M.} \& \bibinfo{author}{Dattani, N.~S.}
\newblock \bibinfo{title}{{Why quantum coherence is not important in the
  Fenna-Matthews-Olsen Complex}}.
\newblock \emph{\bibinfo{journal}{J. Chem. Theory Comput.}}
  \bibinfo{pages}{150304125043002} (\bibinfo{year}{2015}).

\bibitem{Prior2010}
\bibinfo{author}{Prior, J.}, \bibinfo{author}{Chin, A.~W.},
  \bibinfo{author}{Huelga, S.~F.} \& \bibinfo{author}{Plenio, M.~B.}
\newblock \bibinfo{title}{{Efficient simulation of strong system-environment
  interactions}}.
\newblock \emph{\bibinfo{journal}{Phys. Rev. Lett.}}
  \textbf{\bibinfo{volume}{105}}, \bibinfo{pages}{1--4} (\bibinfo{year}{2010}).

\bibitem{Shi2004}
\bibinfo{author}{Shi, Q.} \& \bibinfo{author}{Geva, E.}
\newblock \bibinfo{title}{{A semiclassical generalized quantum master equation
  for an arbitrary system-bath coupling}}.
\newblock \emph{\bibinfo{journal}{J. Chem. Phys.}}
  \textbf{\bibinfo{volume}{120}}, \bibinfo{pages}{12063}
  (\bibinfo{year}{2004}).

\bibitem{Stockburger2002}
\bibinfo{author}{Stockburger, J.~T.} \& \bibinfo{author}{Grabert, H.}
\newblock \bibinfo{title}{{Exact c-number representation of non-Markovian
  quantum dissipation.}}
\newblock \emph{\bibinfo{journal}{Phys. Rev. Lett.}}
  \textbf{\bibinfo{volume}{88}}, \bibinfo{pages}{170407}
  (\bibinfo{year}{2002}).

\bibitem{Xu2005}
\bibinfo{author}{Xu, R.~X.}, \bibinfo{author}{Cui, P.}, \bibinfo{author}{Li,
  X.~Q.}, \bibinfo{author}{Mo, Y.} \& \bibinfo{author}{Yan, Y.}
\newblock \bibinfo{title}{{Exact quantum master equation via the calculus on
  path integrals}}.
\newblock \emph{\bibinfo{journal}{J. Chem. Phys.}}
  \textbf{\bibinfo{volume}{122}}, \bibinfo{pages}{041103}
  (\bibinfo{year}{2005}).

\bibitem{Huo2010}
\bibinfo{author}{Huo, P.}, \bibinfo{author}{Bonella, S.},
  \bibinfo{author}{Chen, L.} \& \bibinfo{author}{Coker, D.~F.}
\newblock \bibinfo{title}{{Linearized approximations for condensed phase
  non-adiabatic dynamics: Multi-layered baths and Brownian dynamics
  implementation}}.
\newblock \emph{\bibinfo{journal}{Chem. Phys.}} \textbf{\bibinfo{volume}{370}},
  \bibinfo{pages}{87--97} (\bibinfo{year}{2010}).

\bibitem{Schroter2014}
\bibinfo{author}{Schr{\"{o}}ter, M.} \emph{et~al.}
\newblock \bibinfo{title}{{Exciton-vibrational coupling in the dynamics and
  spectroscopy of Frenkel excitons in molecular aggregates}}.
\newblock \emph{\bibinfo{journal}{Phys. Rep.}} \textbf{\bibinfo{volume}{567}},
  \bibinfo{pages}{1--78} (\bibinfo{year}{2014}).

\bibitem{Fain2000}
\bibinfo{author}{Fain, B.}
\newblock \emph{\bibinfo{title}{{Irreversibilities in Quantum Mechanics}}}
  (\bibinfo{publisher}{Kluer Acadenic Publishers},
  \bibinfo{address}{Dorderecht}, \bibinfo{year}{2000}).

\bibitem{Jiang1991}
\bibinfo{author}{Jiang, X.-p.} \& \bibinfo{author}{Brumer, P.}
\newblock \bibinfo{title}{{Creation and dynamics of molecular states prepared
  with coherent vs partially coherent pulsed light}}.
\newblock \emph{\bibinfo{journal}{J. Chem. Phys.}}
  \textbf{\bibinfo{volume}{94}}, \bibinfo{pages}{5833--5843}
  (\bibinfo{year}{1991}).

\bibitem{Mancal2010}
\bibinfo{author}{Man{\v{c}}al, T.} \& \bibinfo{author}{Valkunas, L.}
\newblock \bibinfo{title}{{Exciton dynamics in photosynthetic complexes:
  Excitation by coherent and incoherent light}}.
\newblock \emph{\bibinfo{journal}{New J. Phys.}} \textbf{\bibinfo{volume}{12}},
  \bibinfo{pages}{1--12} (\bibinfo{year}{2010}).

\bibitem{Brumer2012a}
\bibinfo{author}{Brumer, P.} \& \bibinfo{author}{Shapiro, M.}
\newblock \bibinfo{title}{{Molecular response in one-photon absorption via
  natural thermal light vs. pulsed laser excitation.}}
\newblock \emph{\bibinfo{journal}{Proc. Natl. Acad. Sci. U. S. A.}}
  \textbf{\bibinfo{volume}{109}}, \bibinfo{pages}{19575--8}
  (\bibinfo{year}{2012}).

\bibitem{Chenu2014}
\bibinfo{author}{Chenu, A.}, \bibinfo{author}{Mal{\'{y}}, P.} \&
  \bibinfo{author}{Man{\v{c}}al, T.}
\newblock \bibinfo{title}{{Dynamic coherence in excitonic molecular complexes
  under various excitation conditions}}.
\newblock \emph{\bibinfo{journal}{Chem. Phys.}} \textbf{\bibinfo{volume}{439}},
  \bibinfo{pages}{100--110} (\bibinfo{year}{2014}).

\bibitem{Mancal2012}
\bibinfo{author}{Man{\v{c}}al, T.} \& \bibinfo{author}{{\v{S}}anda, F.}
\newblock \bibinfo{title}{{Quantum master equations for non-linear optical
  response of molecular systems}}.
\newblock \emph{\bibinfo{journal}{Chem. Phys. Lett.}}
  \textbf{\bibinfo{volume}{530}}, \bibinfo{pages}{140--144}
  (\bibinfo{year}{2012}).

\bibitem{Olsina2012}
\bibinfo{author}{Ol{\v{s}}ina, J.} \& \bibinfo{author}{Man{\v{c}}al, T.}
\newblock \bibinfo{title}{{Parametric projection operator technique for second
  order non-linear response}}.
\newblock \emph{\bibinfo{journal}{Chem. Phys.}} \textbf{\bibinfo{volume}{404}},
  \bibinfo{pages}{103--115} (\bibinfo{year}{2012}).

\bibitem{MukamelBook}
\bibinfo{author}{Mukamel, S.}
\newblock \emph{\bibinfo{title}{{Principles of nonlinear spectroscopy}}}
  (\bibinfo{publisher}{Oxford University Press}, \bibinfo{address}{Oxford},
  \bibinfo{year}{1995}).

\bibitem{Page1952}
\bibinfo{author}{Page, C.~H.}
\newblock \bibinfo{title}{{Instantaneous Power Spectra}}.
\newblock \emph{\bibinfo{journal}{J. Appl. Phys.}}
  \textbf{\bibinfo{volume}{23}}, \bibinfo{pages}{103} (\bibinfo{year}{1952}).

\bibitem{Peterman1997}
\bibinfo{author}{Peterman, E. J.~G.}, \bibinfo{author}{Pullerits, T.},
  \bibinfo{author}{{Van Grondelle}, R.} \& \bibinfo{author}{{Van Amerongen},
  H.}
\newblock \bibinfo{title}{{Electron-phonon coupling and vibronic fine structure
  of light-harvesting complex II of green plants: Temperature dependent
  absorption and high-resolution fluorescence spectroscopy}}.
\newblock \emph{\bibinfo{journal}{J. Phys. Chem. B}}
  \textbf{\bibinfo{volume}{101}}, \bibinfo{pages}{4448--4457}
  (\bibinfo{year}{1997}).

\bibitem{Nordlund1981}
\bibinfo{author}{Nordlund, T.~M.} \& \bibinfo{author}{Knox, W.~H.}
\newblock \bibinfo{title}{{Lifetime of Fluorescence from Light-Harvesting
  Chlorophyll a/b proteins}}.
\newblock \emph{\bibinfo{journal}{Biophys. J.}} \textbf{\bibinfo{volume}{36}},
  \bibinfo{pages}{193--201} (\bibinfo{year}{1981}).

\bibitem{Schodel1998}
\bibinfo{author}{Schodel, R.}, \bibinfo{author}{Irrgang, K.~D.},
  \bibinfo{author}{Voigt, J.} \& \bibinfo{author}{Renger, G.}
\newblock \bibinfo{title}{{Rate of carotenoid triplet formation in solubilized
  light-harvesting complex II (LHCII) from spinach}}.
\newblock \emph{\bibinfo{journal}{Biophys. J.}} \textbf{\bibinfo{volume}{75}},
  \bibinfo{pages}{3143--53} (\bibinfo{year}{1998}).

\bibitem{Duffy2013}
\bibinfo{author}{Duffy, C. D.~P.} \emph{et~al.}
\newblock \bibinfo{title}{{Modeling of fluorescence quenching by lutein in the
  plant light-harvesting complex LHCII.}}
\newblock \emph{\bibinfo{journal}{J. Phys. Chem. B}}
  \textbf{\bibinfo{volume}{117}}, \bibinfo{pages}{10974--86}
  (\bibinfo{year}{2013}).

\bibitem{Polivka2002}
\bibinfo{author}{Pol{\'{\i}}vka, T.}, \bibinfo{author}{Zigmantas, D.} \&
  \bibinfo{author}{Sundstr{\"{o}}m, V.}
\newblock \bibinfo{title}{{Carotenoid S1 State in a Recombinant
  Light-Harvesting Complex of Photosystem II}}.
\newblock \emph{\bibinfo{journal}{Biochemistry}} \textbf{\bibinfo{volume}{41}},
  \bibinfo{pages}{439} (\bibinfo{year}{2002}).

\bibitem{Valkunas2012}
\bibinfo{author}{Valkunas, L.}, \bibinfo{author}{Chmeliov, J.},
  \bibinfo{author}{Kr{\"{u}}ger, T. P.~J.}, \bibinfo{author}{Ilioaia, C.} \&
  \bibinfo{author}{van Grondelle, R.}
\newblock \bibinfo{title}{{How Photosynthetic Proteins Switch}}.
\newblock \emph{\bibinfo{journal}{J. Phys. Chem. Lett.}}
  \textbf{\bibinfo{volume}{3}}, \bibinfo{pages}{2779--2784}
  (\bibinfo{year}{2012}).

\bibitem{Chmeliov2013}
\bibinfo{author}{Chmeliov, J.}, \bibinfo{author}{Valkunas, L.},
  \bibinfo{author}{Kr{\"{u}}ger, T. P.~J.}, \bibinfo{author}{Ilioaia, C.} \&
  \bibinfo{author}{van Grondelle, R.}
\newblock \bibinfo{title}{{Fluorescence blinking of single major
  light-harvesting complexes}}.
\newblock \emph{\bibinfo{journal}{New J. Phys.}} \textbf{\bibinfo{volume}{15}},
  \bibinfo{pages}{85007} (\bibinfo{year}{2013}).

\bibitem{Chmeliov2015b}
\bibinfo{author}{Chmeliov, J.} \emph{et~al.}
\newblock \bibinfo{title}{{An 'all pigment' model of excitation quenching in
  LHCII}}.
\newblock \emph{\bibinfo{journal}{Phys. Chem. Chem. Phys.}}
  \textbf{\bibinfo{volume}{17}}, \bibinfo{pages}{15857--15867}
  (\bibinfo{year}{2015}).

\bibitem{SMSignalsBook}
\bibinfo{editor}{Barkai, E.}, \bibinfo{editor}{Brown, F. L.~H.},
  \bibinfo{editor}{Orrit, M.} \& \bibinfo{editor}{Yang, H.} (eds.)
  \emph{\bibinfo{title}{{Theory and Evaluation of Single-Molecule Signals}}}
  (\bibinfo{publisher}{World Scientific}, \bibinfo{year}{2008}).

\bibitem{VanRoon2000}
\bibinfo{author}{{Van Roon}, H.}, \bibinfo{author}{{Van Breemen}, J. F.~L.},
  \bibinfo{author}{{De Weerd}, F.~L.}, \bibinfo{author}{Dekker, J.~P.} \&
  \bibinfo{author}{Boekema, E.~J.}
\newblock \bibinfo{title}{{Solubilization of green plant thylakoid membranes
  with n-dodecyl-a,D-maltoside. Implications for the structural organization of
  the Photosystem II, Photosystem I, ATP synthase and cytochrome b6f
  complexes}}.
\newblock \emph{\bibinfo{journal}{Photosynth. Res.}}
  \textbf{\bibinfo{volume}{64}}, \bibinfo{pages}{155--166}
  (\bibinfo{year}{2000}).

\bibitem{Nussberger1994}
\bibinfo{author}{Nussberger, S.} \emph{et~al.}
\newblock \bibinfo{title}{{Spectroscopic characterization of three different
  monomeric forms of the main chlorophyll a/b binding protein from chloroplast
  membranes.}}
\newblock \emph{\bibinfo{journal}{Biochemistry}} \textbf{\bibinfo{volume}{33}},
  \bibinfo{pages}{14775--14783} (\bibinfo{year}{1994}).

\bibitem{vanKampen}
\bibinfo{author}{van Kampen, N.~G.}
\newblock \emph{\bibinfo{title}{{Stochastic processes in physics and
  chemistry}}} (\bibinfo{publisher}{North Holland}, \bibinfo{year}{1992}).

\end{thebibliography}
\end{document}